\documentclass[aps,pra,showpacs,twoside,10pt,floatfix,nofootinbib,longbibliography,twocolumn]{revtex4-1}
\usepackage[colorlinks=true, citecolor=red, urlcolor=blue ]{hyperref}
\usepackage{epsfig,newlfont,amssymb,amsfonts,amsmath,bm,subfigure,palatino,mathtools,amsthm,braket,soul,enumitem,color,times,comment,geometry,xcolor}
\usepackage[normalem]{ulem}
\definecolor{iitcolor}{HTML}{F7A600}
\definecolor{checkcolor}{HTML}{7541C0}
\begin{document}

\title{Quantum state transfer using 1D Heisenberg Hamiltonian on quasi-1D lattices}
\author{Chandrima B. Pushpan, Harikrishnan K. J., Amit Kumar Pal}
\affiliation{Department of Physics, Indian Institute of Technology Palakkad, Palakkad 678 623, India}
\date{\today}

\begin{abstract}
We consider transfer of single and multi-qubit states on a quasi-1D lattice, where the time evolutions involved in the state transfer protocol are generated by only 1D Hamiltonians. We use the quasi-1D isotropic Heisenberg model under a magnetic field along the $z$ direction, where the spin-spin interaction strengths along the vertical sublattices, referred to as rungs, are much stronger than the interactions along other sublattices. Tuning the field-strength to a special value, in the strong rung-coupling limit, the quasi-1D isotropic Heisenberg model can be mapped to an effective 1D XXZ model, where each rung mimics an effective two-level system. Consequently, the transfer of low-energy rung states from one rung to another can be represented by a transfer of an arbitrary single-qubit state from one lattice site to another using the 1D XXZ model. Exploiting this, we propose protocols for transferring arbitrary single-qubit states from one lattice site to another by using specific encoding of the single-qubit state into a low-energy rung state, and a subsequent decoding of the transferred state on the receiver rung. These encoding and decoding protocols involve a time evolution generated by the 1D rung Hamiltonian and single-qubit phase gates, ensuring that all time-evolutions required for transferring the single-qubit state are generated from 1D Hamiltonians. We show that the performance of the single-qubit state transfer using the proposed protocol is always better than the same when a time-evolution generated by the full quasi-1D Hamiltonian is used.   
\end{abstract}

\maketitle

\section{Introduction}
\label{sec:intro}

Since the inception of quantum information theoretic protocols~\cite{nielsen2010,*wilde_book}, low-dimensional interacting quantum spin models have served as ideal testing grounds~\cite{Amico2008,*Latorre_2009,*Modi2012,*laflorencie2016,*DeChiara_2018,*Bera_2018}, leading to the growth of a vast interdisciplinary area of research. Paradigmatic utilization of quantum spin models in quantum information science and technology includes quantum state transfer via one-dimensional (1D) quantum spin models~\cite{Bose2003,*Bose2013_chapter}, measurement-based quantum computation using cluster states arising out of Ising interactions between spins~\cite{raussendorf2001,*raussendorf2003,*briegel2009,*Wei2018}, and topological quantum error correction on quantum lattice models of specific geometry ~\cite{kitaev2001,*kitaev2006,*bombin2006,*bombin2007}. Realizations of such low-dimensional quantum spin models using trapped ions~ \cite{Porras2004,*Leibfried2005,*monz2011,*Korenblit_2012,*Bohnet2016}, superconducting qubits~\cite{barends2014,*Yariv2020}, nuclear magnetic resonance~ \cite{Vandersypen2005,*Negrevergne2006}, solid-state systems~\cite{Schechter2008,*Bradley2019}, and ultra-cold atoms~ \cite{Greiner2002,*Duan2003,*Bloch_2005,*Bloch2008,*Struck2013} have also extended the possibility of implementing these quantum protocols beyond the constraints of being only theoretical.

\begin{figure*}
    \centering
    \includegraphics[width=0.7\textwidth]{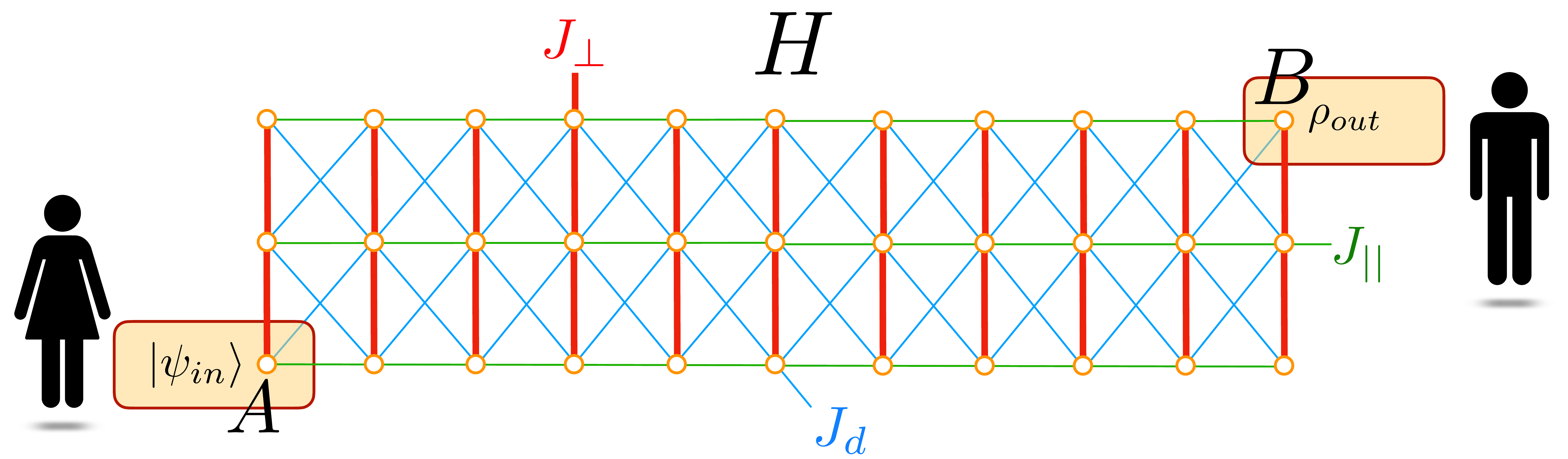}
    \caption{A quasi-1D zig-zag lattice of $11$ rungs and $3$ legs, with one of $11\times 3$ spin-$\frac{1}{2}$ particles on each lattice site. The spin-exchange interaction strength $J_{\perp}$ along the rungs is much stronger than the same along the legs, denoted by  $J_{||}$, and the strength along the diagonals, denoted by  $J_d$. In the state transfer protocol,  Alice initiates a qubit $A$ in the state $\ket{\psi_{in}}$ at $t=0$, and Bob collects the state $\rho_{out}(t^\prime)$ from the qubit $B$ at a pre-decided time $t=t^\prime$, where $\rho_{out}(t^\prime)$ can be a close approximation of the single-qubit state $\ket{\psi_{in}}$. The transfer of the state takes place via the dynamics of the spin system described by the Hamiltonian $H$.}
    \label{fig:basic_state_transfer}
\end{figure*}

Among the quantum protocols utilizing the properties of quantum spin models, quantum state transfer~\cite{Bose2003,*Bose2013_chapter} has arguably been one of the most prominent ones. The protocol aims to send a quantum state $\ket{\psi_{in}}$ of a single or a collection of qubits~\cite{Burgarth2005,*Burgarth2005a,Burgarth2005b,Vaucher_2005}, represented by spin-$1/2$ particles, in possession of Alice, the \emph{sender}, to Bob, the \emph{receiver} (see Fig.~\ref{fig:basic_state_transfer} for a schematic representation of the protocol). The \emph{communication channel} between Alice and Bob is a \emph{low-dimensional} lattice hosting a set of qubits in the state $\ket{\psi_{ch}}$.  Among the channel qubits, Bob has a collection of the same number of qubits as Alice in his possession. The combined system of the sender, the channel, and the receiver, is described by the initial state $\ket{\psi_{in}}\otimes\ket{\psi_{ch}}$, which evolves under the system-Hamiltonian $H$ to a state $\ket{\psi(t)}=\text{e}^{-\text{i}Ht}\left[\ket{\psi_{in}}\otimes\ket{\psi_{ch}}\right]$ at time $t$. The state transfer scheme concludes at a pre-decided time $t=t^\prime$ with Bob collecting the state $\rho_{out}(t^\prime)$ on his qubits, obtained by tracing out all other qubits in the state $\ket{\psi(t^\prime)}$. The quality of this state transfer is assessed by $\rho_{out}(t^\prime)$ having a high fidelity $\langle\psi_{in}|\rho_{out}(t^\prime)|\psi_{in}\rangle$ with the input state $\ket{\psi_{in}}$.

After its introduction using one-dimensional (1D) quantum spin models ~\cite{Bose2003,*Bose2013_chapter}, quantum state transfer protocol has been studied in various setups including using a pair of uncoupled~\cite{Burgarth2005,*Burgarth2005a} and coupled~\cite{Burgarth2005b,Li2005,*Almeida2019} spin chains, as well as using different quantum spin models such as the XX model~\cite{Kay2010,Yao2011,*ACOSTACODEN2021}, XY model~\cite{Christandl2004,Banchi2013,*Karbach2005}, and XXZ model~\cite{Subrahmanyam2004,*LIU2008,*FELDMAN20091719,*Pouyandeh2015,*Yang2015,*Shan2018}, along with quantum spin models having long-range correlations~\cite{ALMEIDA2018,*Hermes2020}. While the transfer of single-qubit quantum states have mostly been reported in literature, transferring entangled states have also been studied~\cite{Ronke2011,*Apollaro2023}. Moreover, alongside the perfect state transfer~\cite{Christandl2004,Kay2010,chapman2015} with $\langle\psi_{in}|\rho_{out}(t^\prime)|\psi_{in}\rangle=1$, the idea of \emph{pretty-good state transfer}~\cite{Sousa2014,*Banchi2017,*Serra2022} with a high value of $\langle\psi_{in}|\rho_{out}(t^\prime)|\psi_{in}\rangle$ has also been put forward. Exploring quantum state transfer with a two-dimensional (2D) lattice model have  been  challenging due to the difficulty in tackling the dynamics of the system for their exponentially increasing Hilbert space dimension with increasing number of spins. Among a variety of 2D models, quasi-1D models~\cite{Ercolessi2003,*Ivanov2009} like quantum spin ladders~\cite{Dagotto1996,*Batchelor2003,*Batchelor2003a,Batchelor2007} with number of lattice sites, $N$, in the horizontal direction being far greater than the number of lattice sites, $L$, in the vertical direction ($N\gg L$) have particularly attracted attention~\cite{Li2005,*Almeida2019}.  There also exist results on the implementation of the protocol on 2D triangular lattices~\cite{Miki2012,*Post2015}.

Existing literature on the quantum state transfer protocol have focused mainly on transferring arbitrary single- or multi-qubit  states without any constraints, and using 1D, or quasi-1D quantum spin models. In this paper, we ask \emph{whether transferring single- and multi-qubit quantum states on a quasi-1D lattice is possible via time-evolutions governed by 1D Hamiltonians only}, and answer this question affirmatively. To formulate the problem, we consider  the Heisenberg model~\cite{Heisenberg1928,*Okwamoto1984,*Aplesnin1999,*Zheng1999,*Costa2003,*Cuccoli2006,*Ju2012,*Verresen2018,*Sariyer2019} in a magnetic field on a quasi-1D rectangular zig-zag lattice of $N\times L$ sites ($N\gg L$), where each lattice site hosts a spin-$1/2$ particle (see Fig.~\ref{fig:basic_state_transfer}). We assume the \emph{strong rung-coupling limit of the system}, i.e.,  the coupling strength $J_{\perp}$ between the pairs of spins along the vertical sublattices, referred to as the \emph{rungs}, is much stronger compared to the coupling strengths between spins along the horizontal sublattices, referred to as the \emph{legs}, as well as the same along the \emph{diagonals}. In this limit, the magnetic field can be set to such a value that a doubly-degenerate ground state on each rung is separated from the higher energy states by an energy gap $\sim J_\perp$ and each rung behaves as an \emph{effective} two-level system. The full system Hamiltonian in the low-energy manifold constituted of the $2^N$ low-lying states can be shown to be represented by a 1D XXZ model~\cite{fisher1964,*giamarchi2004,*Mila_2000,*Franchini2017}  up to leading order in perturbation theory~\cite{totsuka1998,*Tonegawa1998,*Mila1998,*Chaboussant1998,*Tribedi2009,kawano1997,*Tandon1999,Pushpan2023}. This leads to a mapping of the transfer of low-energy states belonging to the ground state manifold of a rung to another rung on to a transfer of an arbitrary single-qubit state to another qubit on a 1D lattice using the time-evolution generated by the  effective 1D XXZ Hamiltonian.

In this paper, we consider this effective 1D transfer of the low-energy rung states through the quasi-1D lattice in detail, and calculate the transfer fidelity as a quantifier for the quality of the state transfer. For a systematic investigation, for a specific set of values of the system parameters in agreement with the perturbation theory, we focus on (a) the maximum transfer fidelity for a given input state over a certain time interval during which the perturbation theory remains valid, and (b) the maximum average transfer fidelity, computed by taking the average of transfer fidelity over a sample of Haar-uniformly generated initial states, and then maximizing it over the time interval during which the perturbation theory holds. We show that although the value of the maximum transfer fidelity as well as the maximum average transfer fidelity  exhibit an overall decrease with increasing transfer distance $r$, quantifying the distance between the input and the receiver rung, one can judiciously choose the system parameters to increase the fidelities. We also propose an appropriate parametrization of rung states having overlap with the high-energy states, and show that the effective 1D transfer works even for rung states that have small overlaps with the high-energy sector.   

Next, we focus on the non-trivial question of transferring single-qubit states from one lattice-site to another on quasi-1D lattice hosting the isotropic Heisenberg Hamiltonian. For periodic boundary condition along the rungs,  we propose specific encoding of the initial states on a rung where one of the qubit is prepared in the state to be transferred, such that the encoded rung state belongs to the low-energy sector, and hence can be sent to another rung via an effective 1D transfer. The transferred rung state on the target rung is then decoded using a specific decoding protocol, such that the desired single-qubit state can be extracted from any of the qubits on the target rung. The encoding and decoding involve an 1D evolution on respectively the input and the output rungs using the rung Hamiltonian, and single-qubit phase gates on specific qubits on the rung. We show that the transfer fidelity of this single-qubit state transfer protocol is equal to the transfer fidelity of the low-energy states on a rung. We also show that the performance of the proposed protocol for single-qubit state transfer using the quasi-1D lattice is always better than the same for transfer  of single-qubit states using a time evolution generated by the quasi-1D Hamiltonian. We further propose a number of possible constructions on the quasi-1D lattices where such effective 1D transfer of single-qubit states can be implemented. 

The rest of the paper is organized as follows. In Sec.~\ref{sec:methodology}, we present the essential details on the mapping of the quasi-1D model to the effective 1D XXZ model, and include the necessary information regarding the two-, three-, and four-leg ladder-like lattices that are used in the rest of the paper. The effective 1D transfer of the low-energy states on a rung from one rung to another is discussed in Sec.~\ref{sec:rung2rung_state_transfer}. In Sec.~\ref{sec:single_qubit_transfer}, we present the protocols involving specific encoding and decoding of rung states for two- and four-leg ladders. Sec.~\ref{sec:outlook} contains the concluding remarks and outlook.

\section{Heisenberg model in strong-rung coupling limit}
\label{sec:methodology}

We now consider the Isotropic Heisenberg model on the quasi-1D zig-zag lattice (see Fig.~\ref{fig:basic_state_transfer}), and discuss its low-energy spectrum in the strong rung-coupling limit. Consider a 2D lattice made of intersecting horizontal and vertical sublattices, called the \emph{legs} and the \emph{rungs}, respectively, where the points of intersection are the lattice sites. Let $L$ and $N$ be the number of lattice sites on a rung and a leg, respectively, such that the 2D lattice is of size $N\times L$. Each of these lattice sites hosts a qubit in the form of a spin-$1/2$ particle, such that the dimension of the Hilbert space for the whole system is $2^{NL}$. We refer to these qubits as the \emph{data} qubits. We work in the limit $N\gg L$, representing a quasi-1D system, examples of which includes the widely studied quantum spin ladders~\cite{Batchelor2007} with two ($L=2$), three ($L=3$), or four ($L=4$) legs, and $N\gg 4$.  The qubits are interacting among each other via isotropic Heisenberg interactions~\cite{fisher1964,*giamarchi2004,*Mila_2000,*Franchini2017}, and an external magnetic field applies to all spins along the $z$ direction. The dimensionless Hamiltonian $H/J_\perp\rightarrow H$ representing the system is given by~\cite{Batchelor2007}
\begin{eqnarray} H&=&\frac{1}{4}\sum_{i=1}^N\sum_{j=1}^{L}\vec{\sigma}_{i,j}.\vec{\sigma}_{i,j+1}\frac{J_{||}}{4J_\perp}\sum_{i=1}^N\sum_{j=1}^L\vec{\sigma}_{i,j}.\vec{\sigma}_{i+1,j}\nonumber\\ &&+\frac{J_{d}}{4J_\perp}\sum_{i=1}^N\sum_{j=1}^{L}\left(\vec{\sigma}_{i,j+1}.\vec{\sigma}_{i+1,j}+\vec{\sigma}_{i,j}.\vec{\sigma}_{i+1,j+1}\right)\nonumber\\ &&-\frac{h}{2J_\perp}\sum_{i=1}^N\sum_{j=1}^L\sigma^z_{i,j}
\label{eq:system_Hamiltonian}
\end{eqnarray}
where $J_\perp$, $J_{||}$, and $J_d$ are the strengths of the spin-exchange interactions along the rungs, legs, and diagonals respectively,  $h$ is the strength of the magnetic field on each lattice site, $\vec{\sigma}_{i,j}\equiv\{\sigma^x_{i,j},\sigma^y_{i,j},\sigma^z_{i,j}\}$ are the standard representation of Pauli operators, and $i$ ($j$) being the rung (leg) indices with $1\leq i\leq N$ ($1\leq j\leq L$), such that the pair $i,j$ represents a lattice site.  In this paper, we consider only antiferromagnetic (AFM) interactions between qubits, setting $J_\perp, J_{||},J_d>0$, although the formalism applies to ferromagnetic (FM) interactions ($J_\perp, J_{||},J_d<0$) also. Note that we have assumed periodic boundary condition (PBC) along both legs and rungs. In the case of open boundary conditions (OBC) along the rungs (legs), $j$ ($i$) runs from $1$ to $L-1$ ($N-1$).  To further simplify the notations, we define the dimensionless constants 
\begin{eqnarray}
    u &=& J_{||}/J_\perp,\quad, v= J_d/J_\perp,\quad  w=h/J_\perp,
\end{eqnarray}
and use $H(u,v,w)$ and $H$ interchangeably. To write Pauli matrices, we use the computational basis $\{\ket{0},\ket{1}\}$, such that $\sigma^z\ket{0}=\ket{0}$, $\sigma^z\ket{1}=-\ket{1}$.     

We now consider the system in the limit $u,v=0$, and recognize that $H(0,0,w)=\sum_{i=1}^N\mathcal{H}_i(w)$, where 
\begin{eqnarray}
    \mathcal{H}_i(w)=\frac{1}{4}\sum_{j=1}^L\vec{\sigma}_{i,j}.\vec{\sigma}_{i,j+1}-\frac{w}{2}\sum_{j=1}^L\sigma^z_{i,j},
    \label{eq:rung_hamiltonian}
\end{eqnarray}
representing a rung $i$ to which a Hilbert space of dimension $2^L$ is associated. We further tune $\omega$ to a value $\omega_c$, such that the ground state of $\mathcal{H}_i(w_c)$ is two-fold degenerate with a ground state energy $E_g$, and the degenerate ground states are represented by $\ket{\mathbf{0}_i}$ and $\ket{\mathbf{1}_i}$, while an energy gap $\sim J_\perp$ exists between the ground states and the states corresponding to higher energy eigenvalues. Therefore, the ground state of the full system is $2^N$-fold degenerate at $w=w_c$, constituting the low energy manifold, and is separated from the HEM by an energy gap $\sim J_\perp$. If the interactions $u$ and $v$ are now turned on, the Hamiltonian of the system can be written as $H=H_0+H^\prime$ with $H_0\equiv H(0,0,w_c)$ and $H^\prime\equiv H(u,v,w-w_c)$. For $u,v,\delta w\ll 1$ where we have defined $\delta w=w-w_c$, $H^\prime$ can be treated as a perturbation due to which
the ground state degeneracy in the system is lifted, while the separation between the low  and the high energy manifold is maintained. Up to first order perturbation, the effective Hamiltonian $H_{\text{eff}}$ corresponding to $H^\prime$ is a 1D Hamiltonian, where a strongly-coupled rung behaves as a single object on each lattice site $i$, $1\leq i\leq N$, and is given by 
\begin{eqnarray}
    H_{\text{eff}} &=& \sum_{l,l^\prime=0}^{2^N-1}\bra{\Psi_{l}}H(u,v,w-w_c)\ket{\Psi_{l^\prime}}\ket{\Psi_{l}} \bra{\Psi_{l^\prime}},
\end{eqnarray}
where $\{\ket{\Psi_l}\}$ are the states in the ground state manifold of $H(0,0,w_c)$. Irrespective of the value of $L$, it can be shown~\cite{Pushpan2023} that $H_{\text{eff}}$ is a 1D XXZ model~\cite{fisher1964,*giamarchi2004,*Mila_2000,*Franchini2017} given by 
\begin{eqnarray}
    H_{\text{eff}} &=& \sum_{i=1}^N\left[ J^{xy}_{\text{eff}}\left(\tau^x_i\tau^x_{i+1}+\tau^y_i\tau^y_{i+1}\right)+J^{zz}_{\text{eff}}\tau^z_i\tau^{z}_{i+1}\right]\nonumber\\ &&+h_{\text{eff}}\sum_{i=1}^N \tau^z_i+h^\prime_{\text{eff}}\left(\tau^z_1+\tau^z_N\right),
\end{eqnarray}
where 
\begin{eqnarray}
\tau^x &=& \ket{\mathbf{0}}\bra{\mathbf{1}}+\ket{\mathbf{1}}\bra{\mathbf{0}} ,\\ 
\label{eq:epauliy}
\tau^y &=& -\text{i}\left[\ket{\mathbf{0}}\bra{\mathbf{1}}-\ket{\mathbf{1}}\bra{\mathbf{0}}\right],\\
\label{eq:epauliz}
\tau^z &=& \ket{\mathbf{0}}\bra{\mathbf{0}}-\ket{\mathbf{1}}\bra{\mathbf{1}}, 
\end{eqnarray}
as long as a value for $w=w_c$ can be found such that the ground states in each rung is two-fold degenerate, so that the rung behaves as an \emph{effective} qubit. Note that the state of an effective qubit is a state of $L$ data qubits on a rung.  The effective coupling constants $J^{xy}_{\text{eff}}$,  $J^{zz}_{\text{eff}}$, $h_{\text{eff}}$, and $h^\prime_{\text{eff}}$ are functions of $u$, $v$, and $\delta w$. In the case of PBC along the legs, $h^\prime_{\text{eff}}=0$.  

The identification of the values of $w_c$ leading to the two-fold ground state degeneracy on a rung is a non-trivial problem, and depends on the boundary condition on the rung. For $L=2$, periodic boundary condition (PBC) is equivalent to the open boundary condition (OBC), and $w_c$ if found to be $w_c=1$, with the doubly degenerate ground states given by 
\begin{eqnarray}
\ket{\mathbf{0}} &=& \ket{00}, 
\ket{\mathbf{1}} = \frac{1}{\sqrt{2}}(\ket{01}-\ket{10}),
\label{eq:standard_basis_2}
\end{eqnarray}
with ground state energy $E_g=-3/4$. The effective coupling constants are given by~\cite{totsuka1998,*Tonegawa1998,*Mila1998,*Chaboussant1998,*Tribedi2009,Pushpan2023}
\begin{eqnarray}
J^{xy}_{\text{eff}} &=& \frac{u-v}{4},  
J^{zz}_{\text{eff}} = \frac{u+v}{8},  
h_{\text{eff}} = \frac{u+v-2\delta w}{4},
\end{eqnarray}
along with
\begin{eqnarray}
h^\prime_{\text{eff}} &=&-(u+v)/8,
\end{eqnarray}
for OBC along the legs. However, for $L>2$, $w$ can be tuned to obtained a doubly degenerate ground state only in the case of OBC for odd $L$, and in both the cases of PBC and OBC for even $L$.  We consider the case of OBC for $L=3$~\cite{kawano1997,*Tandon1999,Pushpan2023}, and obtain $w_c=3/2$, at which the ground state energy is $E_g=-7/4$, with 
\begin{eqnarray}
\ket{\mathbf{0}} &=& \ket{000},   
\ket{\mathbf{1}} =\frac{1}{\sqrt{6}}(\ket{001}-2\ket{010}+\ket{100}). 
\label{eq:standard_basis_3}
\end{eqnarray}
The effective coupling constants in this case are given by 
\begin{eqnarray}
J^{xy}_{\text{eff}} &=& \frac{3u-4v}{12}, 
J^{zz}_{\text{eff}}=\frac{9u+8v}{72},  
h_{\text{eff}} =\frac{9u+22v-18\delta w}{36},\nonumber\\
\end{eqnarray}
along with 
\begin{eqnarray}
    h^\prime_{\text{eff}} &=& -(9u+22v)/72. 
\end{eqnarray}
For $L=4$ and with OBC along the rungs, the doubly-degenerate ground states at $w_c=1+1/\sqrt{2}$ are given by
\begin{eqnarray}
\ket{\mathbf{0}} &=& \ket{0000},\nonumber\\   
\ket{\mathbf{1}} &=&\frac{-\ket{0001}+a\ket{0010}-a\ket{0100}+\ket{1000}}{\sqrt{2+2a^2}},
\label{eq:standard_basis_4}
\end{eqnarray}
with $a=1+\sqrt{2}$. The effective coupling constants in this case are 
\begin{eqnarray}
J^{xy}_{\text{eff}} &=&\frac{4u-(2+3\sqrt{2})v}{16}, 
J^{zz}_{\text{eff}}=\frac{6u+(2\sqrt{2}+5)v}{64}, \nonumber\\
h_{\text{eff}} &=&\frac{10u+(2\sqrt{2}+19)v-16\delta w}{32},\nonumber\\
h_{\text{eff}}^\prime &=&  -\frac{10u+(2\sqrt{2}+19)v}{64}. 
\end{eqnarray}
On the other hand, for PBC along the rungs and for even $L$, $\ket{\mathbf{1}}$ is given by~\cite{Pushpan2023} 
\begin{eqnarray}
    \ket{\mathbf{1}} &=&\frac{1}{\sqrt{L}}\sum_{j=1}^{L}(-1)^j\ket{j},
\end{eqnarray}
where 
\begin{eqnarray}
\ket{j}=\ket{1_j}\otimes_{\underset{j^\prime\neq j}{j^\prime=1}}^L\ket{0_{j^\prime}}.
\label{eq:ketj}
\end{eqnarray}
Explicitly, for a four-qubit rung, 
\begin{eqnarray}
    \ket{\mathbf{0}} &=& \ket{0000},\nonumber\\ 
    \ket{\mathbf{1}} &=&\frac{1}{2}(\ket{0001}-\ket{0010}+\ket{0100}-\ket{1000}),
    \label{eq:four_qubit_rung_states}
\end{eqnarray}
which are degenerate at $w_c=2$, and the effective coupling constants are given by 
\begin{eqnarray}
    J_{\text{eff}}^{xy} &=& \frac{u-2v}{4},\quad J_{\text{eff}}^{zz} = \frac{u+2v}{16},\nonumber\\
    h_{\text{eff}} &=& \frac{3(u+2v)-4\delta w}{8},\quad h_{\text{eff}}^\prime = -\frac{3(u+2v)}{16}.
\end{eqnarray}
See~\cite{Pushpan2023} for a calculation of the effective coupling constants  for arbitrary system-size.

In this paper, we are specifically interested in the ladder geometry of the lattice, and concentrate on the cases $L=2,3$, and $4$ for further discussions. We point out here that a state $\rho$ in the low-energy subspace of the  Hilbert space of $H$, written using the computational basis $\{\ket{0},\ket{1}\}$ of the data qubits, is also  a state of the 1D effective XXZ model, and can be written in terms of the states $\{\ket{\mathbf{0}},\ket{\mathbf{1}}\}$ of the effective qubit. In the later case, we denote the state by $\varrho$, where the state $\rho^\prime=\rho$ can be obtained from $\varrho$ via Eqs.~(\ref{eq:standard_basis_2}), (\ref{eq:standard_basis_3}), and (\ref{eq:standard_basis_4}) in the cases of $L=2,3$, and $4$ respectively. Note also that the \emph{low-energy component} of a generic state $\rho$ in the Hilbert space of the full system can be extracted as $\varrho=P\rho P$, where $P=\sum_{\mathbf{k}}P_{\mathbf{k}}$,  with $P_{\mathbf{k}}=\ket{\mathbf{k}}\bra{\mathbf{k}}$,  $\ket{\mathbf{k}}=\otimes_{i=1}^N \ket{k_i}$, $k_i=\mathbf{0},\mathbf{1}$. One can also obtain the state $\rho^\prime$ from $\varrho$ via Eqs.~(\ref{eq:standard_basis_2}), (\ref{eq:standard_basis_3}), and (\ref{eq:standard_basis_4}) in the case of the spin ladders under focus, while in general, $\rho^\prime\neq \rho$ due to a non-zero high-energy component in $\rho$.

\section{Rung-to-rung state transfer}
\label{sec:rung2rung_state_transfer}

\begin{figure}
    \centering
    \includegraphics[width=0.8\linewidth]{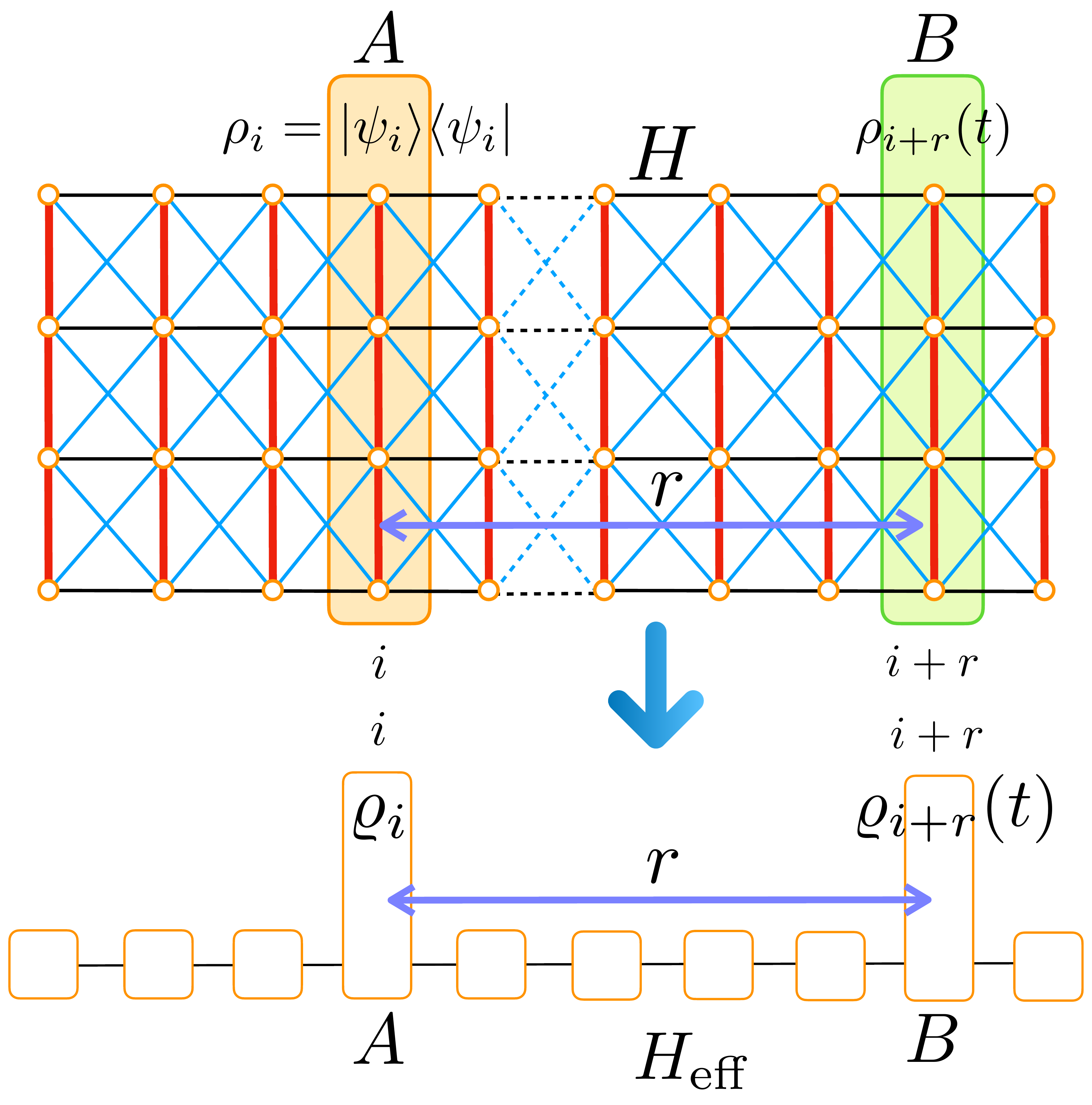}
    \caption{Schematic representation of the transfer of a rung state $\ket{\psi_i}$ through the 2D lattice using the Hamiltonian $H$ (Eq.~(\ref{eq:system_Hamiltonian})) from the rung $i$ to the rung $i+r$, and the corresponding transfer of the low-energy component $\varrho_i$ through the 1D lattice from the site $i$ to site $i+r$ using the low-energy effective Hamiltonian $\tilde{H}$. See Sec.~\ref{sec:rung2rung_state_transfer}.}
    \label{fig:state_transfer_scheme}
\end{figure}

We now consider the scenario where both the sender, Alice, and the receiver, Bob, have a rung each in their possession. Also, for initialization of the system,  access to only the subspace spanned by the ground states of the rungs is available. Under these assumptions, we discuss  the protocol for rung-to-rung (R--R) state transfer through a spin ladder in the strong rung coupling limit. Assuming that the magnetic field can be tuned to get a doubly degenerate ground state on each rung, the most general $L$-qubit state that Alice can transfer through the ladder from, say, rung $i$, to the rung $i+r$ at a distance $r$ is given by
\begin{eqnarray}
    \ket{\psi_i} &=& a_1\ket{\mathbf{0}_i}+\text{e}^{\text{i}a_2}\sqrt{1-a_1^2}\ket{\mathbf{1}_i}.
    \label{eq:rung_initial_state}
\end{eqnarray}
Here,  $\{a_1,a_2\}$ are real numbers with $0\leq |a_1|\leq 1$, and $0\leq a_2<2\pi$, and the forms of $\ket{\mathbf{0}},\ket{\mathbf{1}}$ are given in Sec.~\ref{sec:methodology}. The protocol for the R--R transfer of a state of the form $\ket{\psi_i}$  is as follows (see Fig.~\ref{fig:state_transfer_scheme} for a schematic representation).

\vspace{0.25cm}
\hrule
\vspace{0.25cm}

\noindent\textbf{Protocol for R--R state transfer}

\begin{enumerate}
    \item[\textbf{1:}] \emph{Initialize.} The system is prepared in the initial state $\rho_{in}=\ket{\Psi_{in}}\bra{\Psi_{in}}$, with     \begin{eqnarray}
       \ket{\Psi_{in}} = \ket{\psi_i}\otimes_{j\neq i}\ket{\mathbf{0}_j},
    \end{eqnarray} 
    where Alice's rung $A\equiv i$  is in the state $\ket{\psi_{i}}$, and all other rungs $j$ in the system are in the ground state $\ket{\mathbf{0}_j}=\ket{0}^{\otimes L}$. This can be achieved by tuning the magnetic field to a value $w=w^\prime$ such that the rung $j$ is in the ground state $\ket{\mathbf{0}_j}$.
    
    \item[\textbf{2:}] \emph{Transfer.} The state $\rho_{in}$ is evolved using the Hamiltonian $H(u,v,\delta w)$ with $u,v,\delta w\ll 1$, such that the system is contained within the low-energy manifold of the system described by the Hamiltonian $H=H(0,0,w_c)+H(u,v,\delta w)$. This leads to the time-evolved state $\rho_{out}(t)=T\rho_{in}T^{-1}$, where 
    \begin{eqnarray}
        T &=& \text{e}^{-\text{i}H(u,v,\delta w) t}.
        \label{eq:system_time_evolution}
    \end{eqnarray}
 
    \item[\textbf{3:}] \emph{Extract.} Bob determines the state $\rho_{i+r}(t)$ on the rung $B\equiv i+r$ at a pre-determined time $t$ as \begin{eqnarray}
    \rho_{i+r}(t)=\text{Tr}_{\overline{i+r}}\rho_{out}(t),
    \label{eq:rung_output_state}
    \end{eqnarray} 
    where the partial trace is taken over all data qubits in the system except the data qubits in the rung $i+r$. This is the state transferred to the rung $i+r$ at time $t$ from the rung $i$ with initial state $\ket{\psi_i}$. 
\end{enumerate}

\vspace{0.25cm}
\hrule
\vspace{0.25cm}

\begin{figure}
    \centering
    \includegraphics[width=\linewidth]{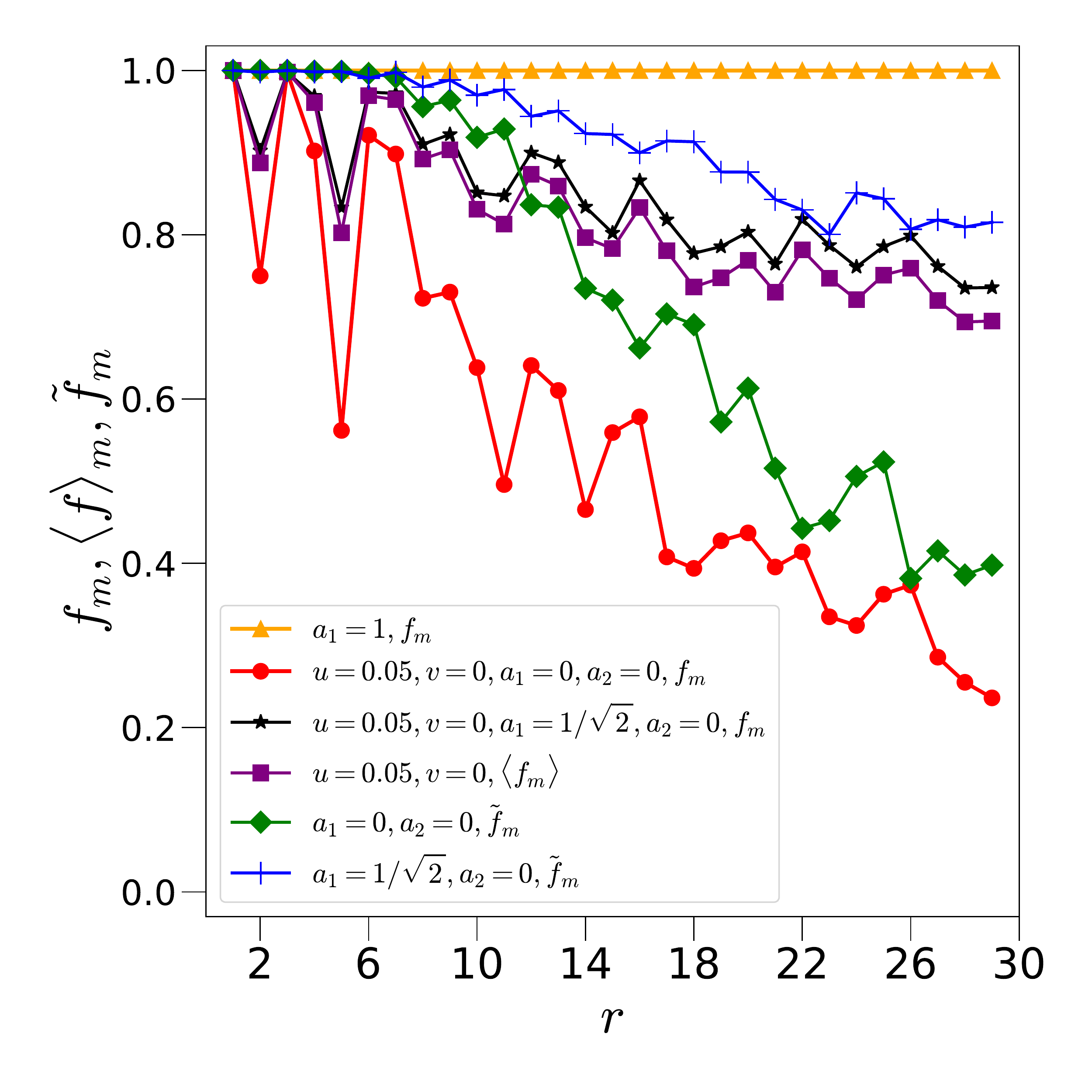}
    \caption{Variations of $f_m$, $\langle f\rangle_m$, and  $\tilde{f}_m$ as functions of transfer distance, $r$, between the sender rung $A\equiv i$ and the receiver rung $B\equiv i+r$, for different initial low-energy rung states of the form (\ref{eq:rung_initial_state}). A quasi-1D lattice of size $30\times 2$ has been used for simulation. For $f_m$ and $\langle f\rangle_m$, we set $u=5\times 10^{-2}, v=0$, while for $\tilde{f}_m$, $u$ and $v$ are set to their optimum values. For all cases, we set $\delta w=0$, and $t\leq J_\perp=10^2$.  All quantities plotted are dimensionless.}
    \label{fig:max_fidelity_r}
\end{figure}

Under the strong-coupling limit, the transfer fidelity (TF) is given by  
\begin{eqnarray}
f=\langle\psi_i|\rho_{i+r}|\psi_i\rangle,
\label{eq:2d_rung_tf}
\end{eqnarray} 
as a function of the initial state parameters and time. We note here that $f_{AB}$ is a function of the initial state parameters $(a_1,a_2)$, time $t$, the transfer distance $r$, and in case of OBC along the legs,  the position of the initial state $i$. However, we refrain from introducing these dependencies in notations to keep the text uncluttered. The maximum TF (MTF) within the allowed range of time is given by 
\begin{eqnarray}
    f_m &=& \max_{t}f. 
\end{eqnarray}
One can also determine the average TF (ATF) as a function of $t$, averaged over a statistically large set of Haar-uniformly sampled initial states of the form $\ket{\psi_{in}}$, given by 
\begin{eqnarray}
    \langle f\rangle &=& \int fP(f)df,
\end{eqnarray}
where the $P(f)$ is the probability distribution of $f$. The maximum of the ATF (MATF) over $t$ is given by $\langle f\rangle_m$. Corresponding to a specific state $\ket{\psi_i}$ on Alice, a \emph{good} state transfer between Alice and Bob is indicated by a high value of $f_m$, while the quality of the overall transfer of the initial state of the form $\ket{\psi_{i}}$ is quantified by $\langle f\rangle_m$. 

Note here that as long as the initial state on the rung $i$ is of the form $\ket{\psi_i}$, the process of the R--R state transfer is \emph{effectively} an 1D process constituted of the following steps.

\vspace{0.25cm}
\hrule
\vspace{0.25cm}

\noindent\textbf{Effective 1D state transfer}

\begin{enumerate}
\item[\textbf{1:}] \emph{Initialize.} The 1D effective system is initialized to 
\begin{eqnarray}
\varrho_{in}=\varrho_i\otimes_{j\neq i}\ket{\mathbf{0}_j}\bra{\mathbf{0}_j},
\end{eqnarray} 
where $\varrho_i=P_{k_{i}}\ket{\psi_i}\bra{\psi_i}P_{k_{i}}$ is the low-energy component of $\ket{\psi_i}$, with $P_{k_i}=\ket{\mathbf{0_i}}\bra{\mathbf{0_i}}+\ket{\mathbf{1}_i}\bra{\mathbf{1}_i}$.  Note that as long as $\ket{\psi_i}$ is of the form (\ref{eq:rung_initial_state}), $\rho^\prime_i$ computed from $\varrho_i$ equals to $\rho_i$ (see Sec.~\ref{sec:methodology}).

\item[\textbf{2:}] \emph{Transfer.} Next, the 1D effective Hamiltonian $H_{\text{eff}}$ is turned on, such that the time-evolved state $\varrho_{out}(t) = T_{\text{eff}}\varrho_{in}T_{\text{eff}}^{-1}$, with
\begin{eqnarray}
      T_{\text{eff}} &=& \text{e}^{-\text{i}H_{\text{eff}} t},
\end{eqnarray} 
is obtained. In order to maintain the perturbation regime, we constraint the time to the range $0\leq t\leq J_\perp$. 

\item[\textbf{3:}] \emph{Extract.} From $\varrho_{out}(t)$, the state
\begin{eqnarray}
    \varrho_{i+r}(t)=\text{Tr}_{\overline{i+r}}\varrho_{out}(t)
\end{eqnarray}
on the effective spin $i+r$ is picked up by tracing out all  effective spins other than the spin $i+r$. 
\end{enumerate} 

\vspace{0.25cm}
\hrule
\vspace{0.25cm}

For $t\leq J_\perp$, the TF corresponding to the ladder is approximated by $f_{\text{eff}}=\langle\phi_i|\varrho_{i+r}|\phi_i\rangle$, where $\varrho_i=\ket{\phi_i}\bra{\phi_i}$ is the low-energy component of $\rho_i=\ket{\psi_i}\bra{\psi_i}$, for all $t\leq J_\perp$. Our numerical analysis suggests that for the family of states~(\ref{eq:rung_initial_state}), the maximum absolute error $\varepsilon=\max_t\left|f-f_{\text{eff}}\right|\leq 10^{-6}$ for all $N\leq 30$, and for all $L\leq 10$.
This implies that the R--R transfer of low-energy states~(\ref{eq:rung_initial_state}) between the rung $i$ and the rung $i+r$ via the isotropic Heisenberg model in the strong rung-coupling limit is faithfully represented by the performance of the 1D XXZ model in transferring a generic single-qubit state $\ket{\psi_i}=a_1\ket{0}+\text{e}^{\text{i}a_2}\sqrt{1-a_1^2}\ket{1}$ from the site $i$ to the site $i+r$, which has earlier been studied in~\cite{Subrahmanyam2004,*LIU2008,*FELDMAN20091719,*Pouyandeh2015,*Yang2015,*Shan2018}. This drastically reduces the complexity of computation of the TF for all rung states of the form (\ref{eq:rung_initial_state}).  Note that this formalism is valid for all rung states~(\ref{eq:rung_initial_state}), for which the \emph{high-energy overlap}, quantified by 
\begin{eqnarray}
    D=1-\sum_{k_i=\mathbf{0},\mathbf{1}}\left|\langle k_i|\psi_i|k_i\rangle\right|^2,
\end{eqnarray}
tends to vanish.


\subsection{Specific examples with \texorpdfstring{$L\leq 4$}{L less than or equal to 4}}

We now discuss the transfer of states of the form (\ref{eq:rung_initial_state}) through two- $(L=2)$ and three-leg $(L=3)$ ladders with OBC along the rungs. We first point out here that irrespective of the value of $L$ and the boundary condition along the rungs,  the state $\ket{\psi_i}$ with $a_1=1,a_2=0$ corresponds to the initial state $\ket{\Psi_{in}}=\ket{0}^{NL}$, which is an eigenstate of the full system Hamiltonian $H$, implying that the TF $f=1$ for all $t$ and for all $r$. However, this is not the case for other states in the family of (\ref{eq:rung_initial_state}). For demonstration, we choose $L=2$ for which a generic state (\ref{eq:rung_initial_state}) in the low-energy sector on the rung $i$ is given by (see Eq.~(\ref{eq:standard_basis_2}))
\begin{eqnarray}
    \ket{\psi_i}=a_1\ket{00}+\text{e}^{\text{i}a_2}\frac{\sqrt{1-a_1^2}}{\sqrt{2}}\left(\ket{01}-\ket{10}\right).
    \label{eq:low_energy_rung_state_L2}
\end{eqnarray}
Our numerical analysis suggests that for all $r\geq 1$, $f$ is invariant with a change in $a_2$, implying $a_1$ to be the only relevant state parameter. The MTFs, $f_m$, for different transfer distances $r$, are plotted in Fig.~\ref{fig:max_fidelity_r} for two different initial states -- (i) the maximally-entangled Bell state $\ket{\psi_i}=(\ket{01}-\ket{10})/\sqrt{2}$ ($a_1=a_2=0$), and (ii) the state $\ket{\psi_i}=[\ket{00}+(\ket{01}-\ket{10})/\sqrt{2}]/\sqrt{2}$ ($a_1=1/\sqrt{2},a_2=0$), where we have used a $30\times 2$ lattice for R--R transfer from $A\equiv i=1$ to $B\equiv i+r$ with $1\leq r\leq 29$, and $t\leq J_\perp=10^2$ to maintain the perturbation regime (see Sec.~\ref{sec:methodology}).  Observe that $f_m$ experiences an overall decrease with increasing $r$ for both initial states, which is also qualitatively true for all states of the family~(\ref{eq:rung_initial_state}), and for all points in the parameter space $(u,v,\delta w)$. We also investigate the behaviour of MATF, $\langle f\rangle_m$, as a function of $r$ in the case of the ladder-like lattice with $L=2,3,4$, where a sample of $10^5$ Haar-uniformly generated initial states of the form~(\ref{eq:rung_initial_state}) is used to determine $\langle f\rangle_m$. The variation of $\langle f\rangle_m$ as a function of $r$ is qualitatively similar to that of the MTF, as shown for the case of $L=2$ in Fig.~\ref{fig:max_fidelity_r}.   Following the same approach, (a) in the case of the three-leg ladder ($L=3$) with OBC along the rungs, we consider the low-energy states on a rung with the form (see Eq.~(\ref{eq:standard_basis_3})) 
\begin{eqnarray}
    \ket{\psi_i} &=& a_1\ket{000}+\frac{\text{e}^{\text{i}a_2}\sqrt{1-a_1^2}}{\sqrt{6}}\left(\ket{001}-2\ket{010}+\ket{100}\right),\nonumber\\ 
    \label{eq:input_3_leg}
\end{eqnarray}
while (b) for the four-leg ladder with OBC along the rungs, the low-energy states on a rung are (see Eq.~(\ref{eq:standard_basis_4})) 
\begin{eqnarray}
    \ket{\psi_i} &=& a_1\ket{0000}+\frac{\text{e}^{\text{i}a_2}\sqrt{1-a_1^2}}{\sqrt{2+2a^2}}\Big(\ket{1000}-a\ket{0100}\nonumber\\ &&+a\ket{0010}-\ket{0001}\Big), 
    \label{eq:input_4_leg}
\end{eqnarray}
with $a=1+\sqrt{2}$. Results on the MTF and MATF regarding the transfer of an initial three- and four-qubit states of respectively the forms~(\ref{eq:input_3_leg}) and ~(\ref{eq:input_4_leg}) are qualitatively similar to the same for the $L=2$ case. We also consider the case of PBC along the rungs, and note that (a) double degeneracy of ground state is obtained only if $L$ is even,  and(b) the PBC and OBC along the rungs in the case of $L=2$ are equivalent. For $L=4$ and PBC along the rungs, 
\begin{eqnarray}
    \ket{\psi_i} &=& a_1\ket{0000}+\frac{\text{e}^{\text{i}a_2}\sqrt{1-a_1^2}}{2}\Big(\ket{0001}-\ket{0010}\nonumber\\ &&+\ket{0100}-\ket{100}\Big). 
    \label{eq:input_4_leg_pbc}
\end{eqnarray}
In all these cases, variations of $f_m$ and $\langle f\rangle_m$ as functions of $r$ are qualitatively similar to the case of $L=2$.

\paragraph{Dependence on system parameters.} So far, we have presented performance of the transfer protocols for multi-qubit low-energy entangled state on a chosen rung via a quasi-1D isotropic Heisenberg model in the strong rung-coupling limit, described by a point in the space of the system parameters $(u,v,\delta w)$ in perturbation regime. The question as to whether an improvement in the performance can be obtained by tuning the system parameters is natural at this point. To answer this, we optimize $f_m$ over the parameter subspace ($0\leq u\leq 10^{-1}$, $0\leq v\leq 10^{-1}$, $0\leq \delta w\leq 10^{-1}$) in the perturbation regime for the states (i) $a_1=a_2=0$, and (ii) $a_1=1/\sqrt{2},a_2=0$, and find that for small values of $r$, $f_m$ can be pushed to unity, or close to unity by a judicious choice of the system parameters $(u,v,\delta w)$. In Fig.~\ref{fig:max_fidelity_r}, we plot the optimized $f_m$, which we denote by $\tilde{f}_m$, as a function of $r$ for a ladder with $L=2$. Our data suggests that $\tilde{f}_m\geq f_m$ for all values of $r\leq 29$ in the case of the rung states $(a_1=a_2=0)$, and $(a_1=1/\sqrt{2},a_2=0)$ for two-leg ladders. Qualitatively similar results are obtained for the case of $L=3$ and $4$ also.

\begin{figure}
    \includegraphics[width=0.8\linewidth]{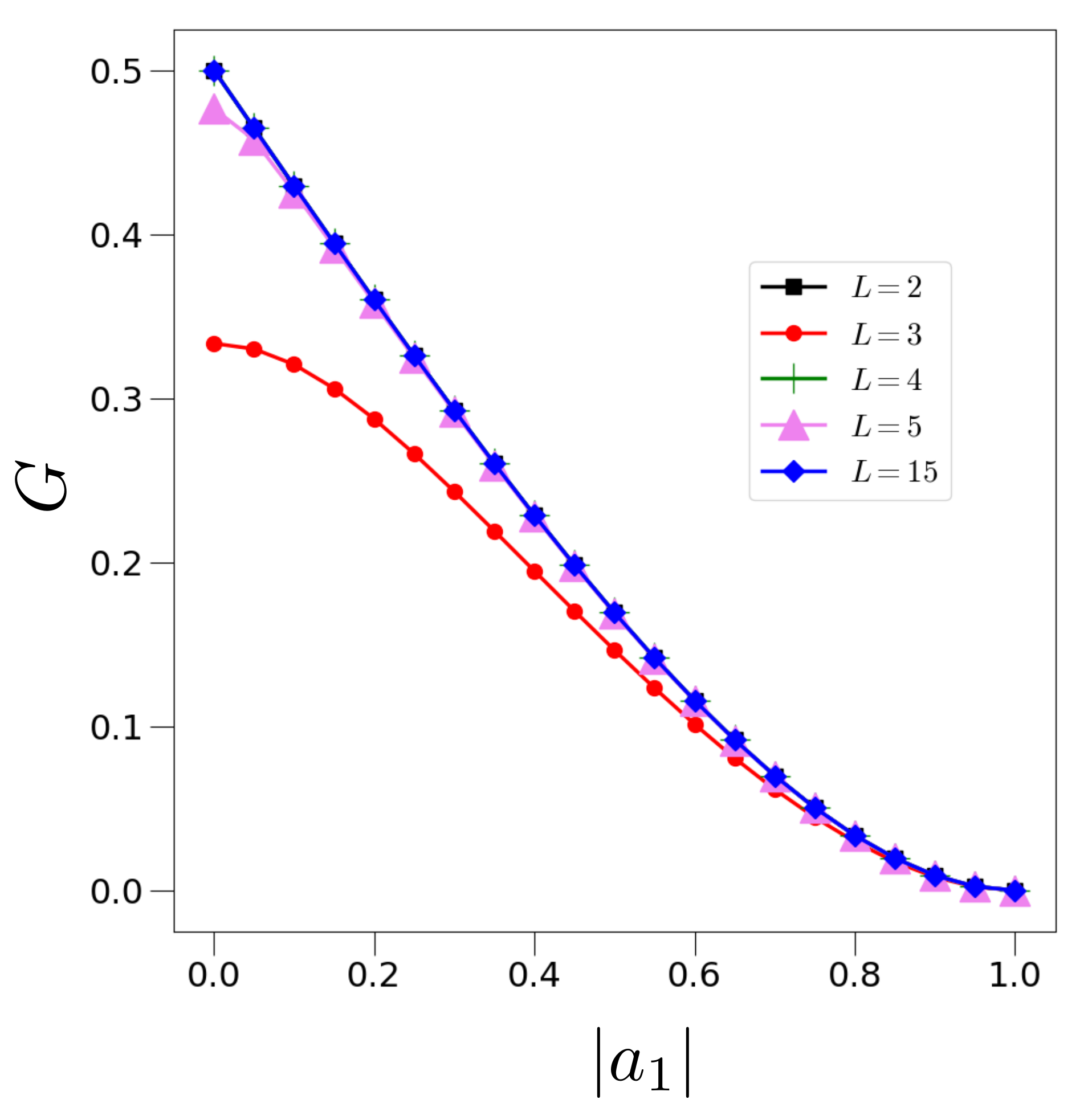}
    \caption{Variation of $G$ for the $L$-qubit states of the form~(\ref{eq:rung_initial_state}) is plotted as a function of $|a_1|$ for different values of $L$. All quantities plotted are dimensionless.}
    \label{fig:ggm}
\end{figure}

\paragraph{Entangled state transfer.} A comment on the entanglement of the low-energy quantum states in the ground state manifold on a rung using the isotropic Heisenberg Hamiltonian on ladder-like lattices  is in order here. The methodology treating the quantum state transfer via a ladder Hamiltonian in the perturbation regime as a transfer of arbitrary single-qubit states through an effective 1D lattice using the XXZ model remains unaltered even with increasing $L$~\cite{Pushpan2023} in the case of both PBC (for even $L$) and OBC along the rungs.  The states of the form (\ref{eq:rung_initial_state}) are genuinely multiparty entangled states for all even and odd $L$ with OBC along the rungs, and the variations of their entanglement content, as quantified by the generalized geometric measure~\cite{Sende2010,*Sadhukhan2017} (see~\cite{ggm_definition} for the definition), $G$, as a function of $|a_1|$ becomes $L$-invariant for $L\geq 5$ (see Fig.~\ref{fig:ggm}. On the other hand, for PBC along rungs with even $L$, the low-energy states of the form  
\begin{eqnarray}
    \ket{\psi_i}&=&a_1\ket{0}^{\otimes L}+\text{e}^{\text{i}a_2}\sqrt{\frac{1-a_1^2}{L}}\sum_{j=1}^L(-1)^j\ket{j}
\end{eqnarray}
with $\ket{j}$ given in Eq.~(\ref{eq:ketj}), are also genuinely multiparty entangled, with the variations of $G$ as a function of $|a_1|$ being almost invariant with a change in $L$. Therefore, from the viewpoint of transferring entangled state using isotropic Heisenberg Hamiltonian with strong rung coupling on a ladder-like lattice, it is sufficient to confine the investigations for lattices up to $L=4$.

\begin{figure*}
    \includegraphics[width=\textwidth]{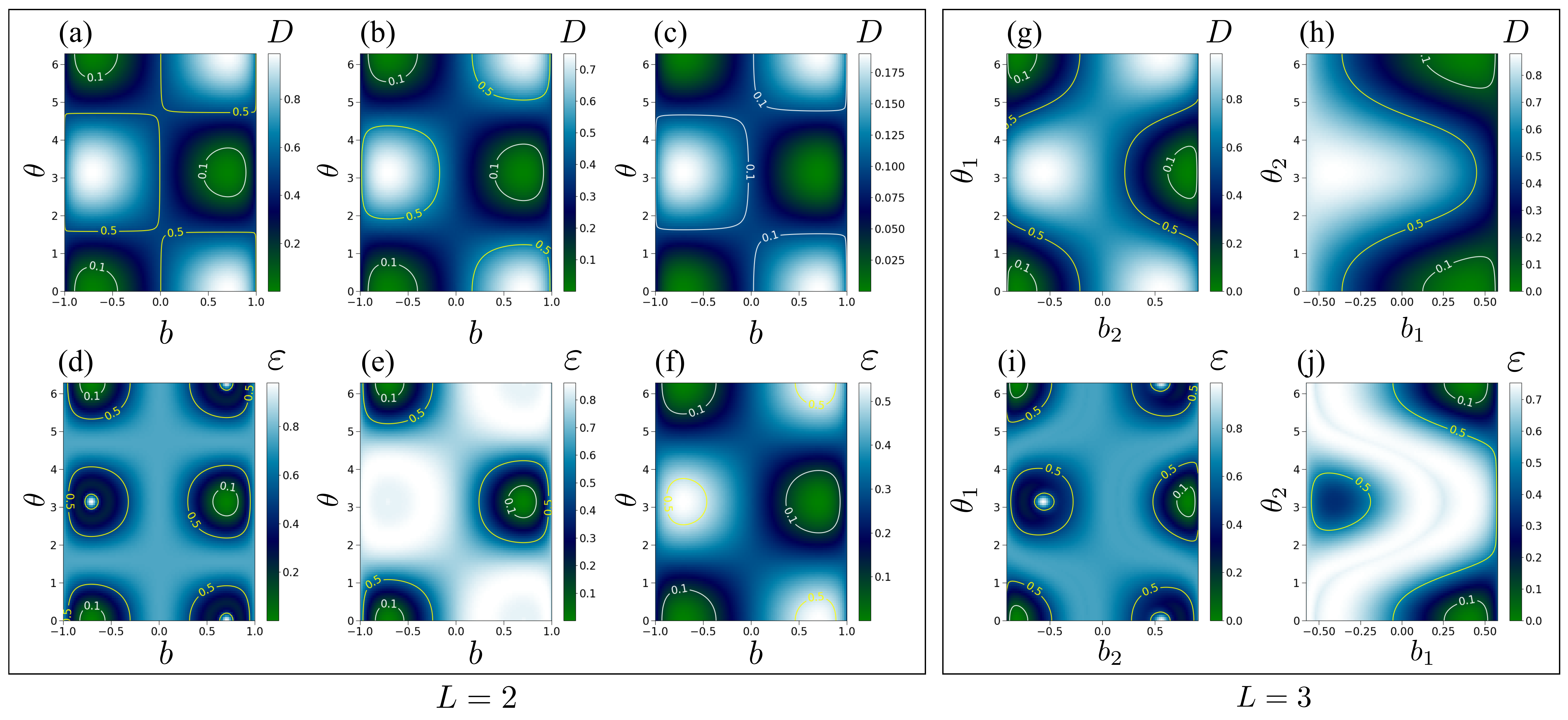}
    \caption{Variations of (a)-(c) $D$ and (d)-(f) $\varepsilon$ as functions of $b$ and $\theta$ for the state (\ref{eq:input_state_high_energy}) in the case of a two-leg ladder for three different values of $a_1=0.1,0.5,0.9$ respectively. Similar variations of (g)-(h) $D$ and (i)-(j) $\varepsilon$ as functions of (g)-(i) $(b_2,\theta_1)$ for $b_1=1/\sqrt{6}$, $\theta_2=0$, and as functions of (h)-(j) $(b_1,\theta_2)$ for $b_2=2/\sqrt{6}$, $\theta_1=\pi$ in the case of a three-leg ladder is shown in figures (g)-(j) in both cases $a_1=0.1$ and $a_2=0$. All quantities plotted are dimensionless, except $\theta$, $\theta_1$, and $\theta_2$, which are in radian.}
    \label{fig:high_energy}
\end{figure*}

\subsection{Transferring high-energy states}

A logical question at this point is whether the transfer of a state $\ket{\psi_i}$ in Alice's possession with non-zero high-energy component is possible via the isotropic Heisenberg model on a ladder-like lattice in the strong rung coupling limit. For investigating this in a  systematic way, we focus on the $L=2$ case and consider the state
\begin{eqnarray}
    \label{eq:input_state_high_energy}
    \ket{\xi_i}=a_1\ket{00}+\text{e}^{\text{i}a_2}a_3\ket{01}+\text{e}^{\text{i}a_4}\sqrt{1-a_1^2-a_3^2}\ket{10},\quad
\end{eqnarray}
where $\{a_1,a_2,a_3,a_4\}$ are real parameters with $0\leq |a_1|,|a_3|\leq 1$, and  $0\leq a_2,a_4<2\pi$. Note that for $L=2$, the state $\ket{11}$  is an eigenstate of the rung $i$ with the highest energy eigenvalue. Therefore its inclusion in writing $\ket{\xi_i}$ only increases the high-energy component of $\ket{\xi_i}$, and hence is discarded.  The high-energy component of $\ket{\xi_i}$ is quantified by $D=|\langle\xi_i^\prime|\xi_i\rangle|^2$, where $\xi^\prime_i=(\ket{01}+\ket{10})/\sqrt{2}$ is the other high energy eigenstate of the rung $i$ apart from $\ket{11}$. Note further that setting $b^2=a_3^2/(1-a_1^2)$ and $\theta=(a_4-a_2)$, one can identify $\ket{\xi_i}\equiv\ket{\psi_i}$ for $b=\pm1/\sqrt{2}$ and $\theta=\pi$. In Fig.~\ref{fig:high_energy}, we plot $D$ (Fig.~\ref{fig:high_energy}(a)-(c)) and   $\varepsilon$ (Fig.~\ref{fig:high_energy}(d)-(f)) as a function of $b$ and $\theta$, where for computing $\varepsilon$, we have fixed $r=2$, and have used an $L=2$-ladder with $N=3$. It is evident from the figures that  for considerable regions of the $(b,\theta)$-parameter space around the points $b=\pm1/\sqrt{2}$, $\theta=\pi$ both the high-energy overlap $D$ and the maximum absolute error $\varepsilon$ is considerably small, but non-zero, implying the existence of subspaces in the parameter space where the rung state transfer is mimicked by the effective 1D XXZ model even when the rung states have a high energy overlap.

In the case of $L=3$, the state $\ket{\psi_i}$ represents a subset of the set of all possible three-qubit W class states, given by 
\begin{eqnarray}
\ket{W_i} &=& a_1\ket{000}+\text{e}^{\text{i}a_2}a_3\ket{001}+\text{e}^{\text{i}a_4}a_5\ket{010}\nonumber\\
&&+\text{e}^{\text{i}a_6}\sqrt{1-a_1^2-a_3^2-a_5^2}\ket{100},
\label{eq:3_leg_state}
\end{eqnarray}
where $\{a_l;l=1,2,\cdots,6\}$ are real, with $0\leq |a_l|\leq 1$ for $l=1,3,5$, and $0\leq a_l<2\pi$ for $l=2,4,6$. Similar to the case of $L=2$, we define 
\begin{eqnarray}
b_1^2=a_3^2/(1-a_1^2),\quad b_2^2=a_5^2/(1-a_1^2),
\end{eqnarray}
and 
\begin{eqnarray}
    \theta_1=a_4-a_2,\quad \theta_2=a_6-a_2,
\end{eqnarray}
where $0\leq |b_1|,|b_2|\leq 1$ and $0\leq \theta_1,\theta_2<2\pi$. In Figs.~\ref{fig:high_energy}(g) and (i), we plot $D$ and $\varepsilon$ respectively as functions of $b_2$ and $\theta_1$, while the same as functions of $b_1$ and $\theta_2$ are plotted in Figs.~\ref{fig:high_energy}(h) and (j) respectively. As is evident from the figures, regions with small non-zero $D$ and $\varepsilon$ exist in the parameter space of $(b_1,b_2,\theta_1,\theta_2)$, similar to the case of $L=2$.  Therefore, there exists states of the form~(\ref{eq:3_leg_state}) with $D\neq 0$ the transfer of which is still effectively a 1D state transfer via a 1D XXZ model. However, this feature quickly disappears as $D$ increases.

\section{Transferring single-qubit states}
\label{sec:single_qubit_transfer}

So far, we have discussed the types of entangled states on a rung that can be transferred R--R using effective 1D quantum state transfer. A question that logically arises at this point is whether an arbitrary single-qubit state   
\begin{eqnarray}
    \ket{\psi_{i,j}}&=&c_0\ket{0}+c_1\ket{1}
    \label{eq:single_qubit_states}
\end{eqnarray}
on a \emph{given} data qubit $(i,j)$ in Alice's possession on the quasi-1D lattice can be transferred using the isotropic Heisenberg Hamiltonian in strong rung-coupling limit. Here, $c_0$ and $c_1$ are complex parameters with $|c_0|^2+|c_1|^2=1$. In this paper, we investigate this question by considering PBC along the rungs, and OBC along the legs. In order to ensure doubly degenerate ground states at $w=w_c$, we consider only cases with even $L$. Note that the energy-constrained R--R transfer protocol works only for rung states $\ket{\psi_i}$ of the form (\ref{eq:rung_initial_state}) with $D=0$, which are in the ground-state subspace of the rung Hamiltonian. A major hurdle in the enterprise of sending single-qubit states is the fact that  the initialization of an arbitrary qubit in a state $\ket{\psi_{i,j}}$  results in the initialization of the rung $i$ in the state
\begin{eqnarray}
\ket{\psi_i}&=&\ket{\psi_{i,j}}\otimes_{j^\prime\neq j} \ket{0_{j^\prime}}=c_0\ket{\phi_0}+c_1\ket{\phi_1},
\label{eq:single_qubit_rung_state}
\end{eqnarray}
with $D\neq 0$,  where  
\begin{eqnarray} 
\ket{\phi_0}&=&\otimes_{j^\prime=1}^L\ket{0_{j^\prime}},\quad \ket{\phi_1}=\ket{1_j}\otimes_{\underset{j^\prime\neq j}{j^\prime=1}}^{L}\ket{0_{j^\prime}}. 
\end{eqnarray} 
We address this challenge with proposals for specific \emph{encoding} of the input state via unitary operations $\mathcal{U}_e\ket{\psi_i}=\ket{\psi_g}$, where  $\ket{\psi_g}$ is a rung state from the ground-state subspace.  However, designing such encoding unitaries for ladders with arbitrary $L$ using a time-evolution generated by the rung Hamiltonian $\mathcal{H}(w_c)$ alone is not possible. Note that an evolution of $\ket{\psi_i}$ due to turning on the rung Hamiltonian $\mathcal{H}_i(w_c)$ would result in
\begin{eqnarray}
    \ket{\psi(t)}&=&\text{e}^{-\text{i}\mathcal{H}_i(w_c)t}\ket{\psi_i},\nonumber\\ 
    &=& \ket{\psi_g}+\sum_{k=\mathbf{2}}^{\mathbf{2^L-1}}\langle k|\psi_i\rangle\text{e}^{-\text{i}(E_k-E_0)t}\ket{k},
    \label{eq:rung_evolution_time}
\end{eqnarray}
upto a global phase, where  
\begin{eqnarray}
\ket{\psi_g}=\langle\mathbf{0}|\psi_i\rangle\ket{\mathbf{0}}+\langle\mathbf{1}|\psi_i\rangle\ket{\mathbf{1}}
\end{eqnarray}
is a state of the form~(\ref{eq:rung_initial_state}), and 
$\ket{k},k=\mathbf{0},\mathbf{1},\cdots,\mathbf{2^L-1}\}$ being the eigenvectors of $\mathcal{H}_i(w_c)$, i.e., 
\begin{eqnarray}
    \mathcal{H}_i(w_c)\ket{k}=E_k\ket{k},
\end{eqnarray}
with $E_k$ as the corresponding eigenvalues, $E_{\mathbf{0}}= E_{\mathbf{1}}\leq \cdots\leq E_{\mathbf{2^L-1}}$. The state $\ket{\Psi(t)}$ has a non-zero high-energy overlap at all $t$,  leading to arriving at the ground state via time evolution impossible. In what follows, we propose two specific examples of such encoding in the cases of the two- and the four-leg ladders ($L=2,4$), where single-qubit phase gates are used along with the time-evolution due to the rung Hamiltonian.

\subsection{Single-qubit state transfer in two-leg ladder} 

We start our discussion with the two-leg ladder, and assume that Alice has the data qubit $(1,j)$ in her possession, where the index of the data qubit is given by $(i,j)$.  The steps of sending an arbitrary single-qubit state via a two-leg ladder are as follows.

\begin{figure*}
    \includegraphics[width=0.7\textwidth]{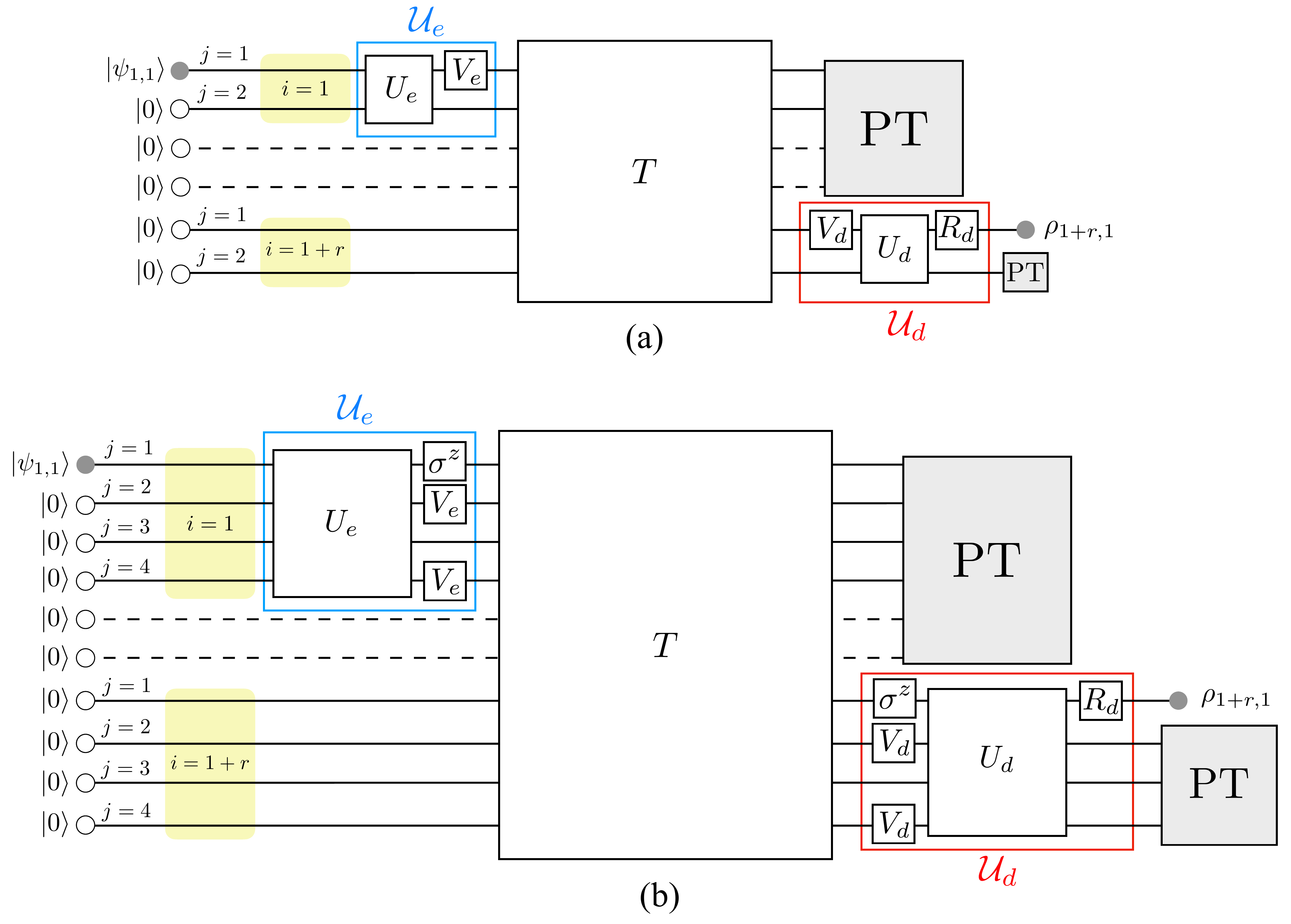}
    \caption{Schematic representation of the protocol for transferring arbitrary single-qubit state using isotropic Heisenberg Hamiltonian in the strong rung-coupling limit on two- and four-leg ladders, where PBC is assumed along the rungs. Partial trace on a set of qubits is represented with a block labelled as ``PT", while the definitions of the unitary operations $T$, $U_e$, $V_e$, $U_d$, $V_d$, and $R_d$ are given in the protocols for the two- and four-leg ladders. For the schematics, it is assumed that Alice has access to the qubit $(1,1)$, while Bob extracts the state at the qubit $(1+r,1)$.}
    \label{fig:circuit}
\end{figure*}

\vspace{0.25cm}
\hrule 
\vspace{0.25cm}

\noindent\textbf{Protocol for two-leg ladder}  

\begin{enumerate}
    \item[\textbf{1:}] \emph{Initialize.} Initialise Alice's qubit on the rung $1$ in the state $\ket{\psi_{1,j}}$ of the form~(\ref{eq:single_qubit_states}). Also, initialize all other qubits in the system to the state $\ket{0}$, such that the state on rung $1$ is 
    \begin{eqnarray}
    \ket{\psi_1} &=& c_0\ket{0_{1,j}0_{1,j^\prime}}+c_1\ket{1_{1,j}0_{1,j^\prime}},  
    \end{eqnarray}
    where $j^\prime=2 (1)$ if $j=1 (2)$. The initial state of the system at $t=0$ is 
    \begin{eqnarray}
        \ket{\Psi_{in}} &=& \ket{\psi_1}\otimes_{i=2}^N\ket{\mathbf{0}_i},
        \label{eq:system_state_input}
    \end{eqnarray}
    where $\ket{\mathbf{0}_{i}}=\ket{0_{i,1}0_{i,2}}$. From now on, we refrain from writing the indices of individual qubits in a rung to keep the text uncluttered.  

    \item[\textbf{2:}] \emph{Encode.} The encoding of the initial state on the rung $1$ into the form~(\ref{eq:low_energy_rung_state_L2}) is done in two steps:  
    \begin{enumerate} 
           \item[\textbf{(a)}] Evolve $\ket{\psi_{1}}$ upto $t=\pi/2$ using the rung Hamiltonian $\mathcal{H}_1(w_c)$ (see Eq.~(\ref{eq:rung_hamiltonian})) such that $\ket{\psi_1}\rightarrow U_e^{(1)}\ket{\psi_1}$ with 
           \begin{eqnarray}
               U_e^{(1)} &=& \text{e}^{-\text{i}\pi\mathcal{H}_1(w_c)/2}.
               \label{eq:rung_evolution}
           \end{eqnarray}
           The explicit form of the time-evolved rung state is 
           \begin{eqnarray}
              \ket{\psi_1} &=& c_0\ket{00}+\frac{c_1\text{e}^{5\text{i}\pi/4}}{\sqrt{2}}(\ket{01}+\text{i}\ket{10})
           \end{eqnarray}
           up to a global phase $\text{e}^{3 \text{i}\pi/8}$.
          
           \item[\textbf{(b)}] Apply the single-qubit unitary operator 
           \begin{eqnarray} 
              V_e^{(1,j)}=\text{e}^{-\text{i}\pi\sigma^z_{(1,j)}/4},
              \label{eq:encoding_unitary_single_qubit}
           \end{eqnarray} 
           on the data qubit $(1,j)$ to transform the state $\ket{\psi_1}\rightarrow V_e^{(1,j)}\ket{\psi_1}$, with the  transformed state 
           \begin{eqnarray}
                 \ket{\psi_1} &=&  c_0\ket{00}+\frac{c_1\text{e}^{5\text{i}\pi/4}}{\sqrt{2}}(\ket{01}-\ket{10}). 
            \end{eqnarray} 
    \end{enumerate}       
    Note that the \emph{encoded} $\ket{\psi_1}$ is now a low-energy state of the two-qubit rung having the form~(\ref{eq:low_energy_rung_state_L2}), while the state of the whole system is of the form~(\ref{eq:system_state_input}). The encoding unitary on rung $1$ is given by 
    \begin{eqnarray}
        \mathcal{U}_e^{(1)}=V_e^{(1,j)}U_e^{(1)}.
        \label{eq:full_encoding_unitary}
    \end{eqnarray}

    \item[\textbf{3:}] \emph{Transfer.} The Hamiltonian $H(u,v,\delta w)$ is turned on with $u,v,\delta w\ll 1$ to evolve the state  using the unitary operator $T$ (see Eq.~(\ref{eq:system_time_evolution})) up to a pre-determined time $t$ such that the system is in the state $\ket{\Psi}=T\ket{\Psi_{in}}$. This is an effective 1D state transfer. See R--R transfer protocol in Sec.~\ref{sec:rung2rung_state_transfer} for details. 

    \item[\textbf{4:}] \emph{Decode.} The decoding of the state on the rung $1+r$ is a three-step process, which depends on the qubit at which Bob intends to extract the state. The steps are as follows. 
    \begin{enumerate} 
           \item[\textbf{(a)}] Given that Alice has access to the qubit $(1,j)$, apply the unitary operator
           \begin{eqnarray} 
              V_d^{(1+r,j)} &=& \text{e}^{\text{i}\pi\sigma^z_{(1+r,j)}/4},
              \label{eq:decoding_unitary_single_qubit}
           \end{eqnarray} 
           on the data qubit $(1+r,j)$ on the rung $1+r$, such that $\ket{\Psi(t)}\rightarrow V_d^{(1+r,j)}\ket{\Psi}$.  Here $j$ is the leg-index of the qubit from which Bob wishes to extract the state.

           \item[\textbf{(b)}] Turn on the rung Hamiltonian $\mathcal{H}(w_c)$ on the rung  $1+r$ to evolve the state $\ket{\Psi}\rightarrow U_d^{(1+r)}\ket{\Psi}$, with $U_d^{(1+r)}$ given by
           \begin{eqnarray}
               U_d^{(1+r)} &=& \text{e}^{-3\text{i}\pi\mathcal{H}_{1+r}(w_c)/2}.
               \label{eq:decoding_rung_unitary}
           \end{eqnarray}

           \item[\textbf{(c)}] Next, apply a local rotation $R_d^{(1+r,j)}$ on the data qubit $(1+r,j)$, such that $\ket{\Psi}\rightarrow R_d^{(1+r,j)}\ket{\Psi}$. The form of $R_d^{(1+r,j)}$ is given by  
           \begin{eqnarray}
           R_d^{(1+r,j)} &=&\left\{ \begin{array}{cc}
              I_{(1+r,j)}, & j=1 \\
              \sigma^z_{(1+r,j)} , & j=2
           \end{array}\right.
           \end{eqnarray}
    \end{enumerate} 
    The full decoding unitary on the rung $(1+r)$ is therefore given by 
    \begin{eqnarray}
        \mathcal{U}_d^{(1+r)}=R_d^{(1+r,j)}U_d^{(1+r)}V_d^{(1+r,j)}.
        \label{eq:full_decoding_unitary}
    \end{eqnarray}

    \item[\textbf{5:}] \emph{Extract.} Bob determines the state $\rho_{1+r,j}$ on the qubit $(1+r,j)$, $j=1,2$, as 
    \begin{eqnarray}
        \rho_{1+r,j}=\text{Tr}_{\overline{1+r,j}}\ket{\Psi}\bra{\Psi}
    \end{eqnarray}
    via tracing out all other qubits in the system except the qubit $(1+r,j)$. 
\end{enumerate}

\vspace{0.25cm}
\hrule 
\vspace{0.25cm}

\begin{figure}
    \centering
    \includegraphics[width=\linewidth]{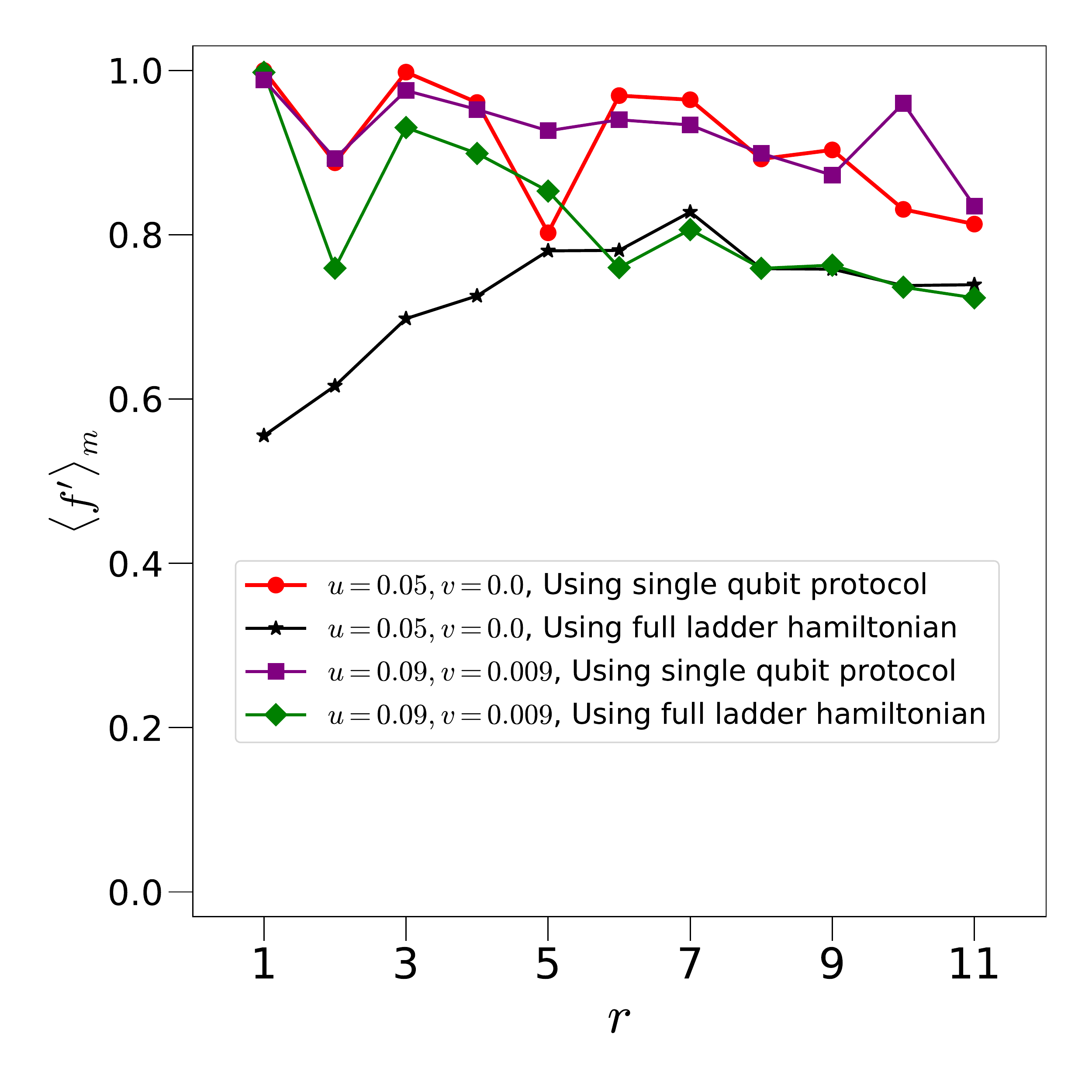}
    \caption{Variations of $\langle f^\prime\rangle_m$ as a function of $r$ for single-qubit state transfer using the proposed protocol and the full quasi-1D Heisenberg Hamiltonian for different values of systems parameters $(u,v)$ with $\delta w=0$ in the perturbation regime. All quantities plotted are dimensionless.}
    \label{fig:single_qubit}
\end{figure}

\noindent A schematic representation of this protocol is depicted in Fig.~\ref{fig:circuit}(a). It is worthwhile to note the following features of the above protocol.

\begin{enumerate}
\item Instead of $U_d^{(1+r)}$, the protocol can also be designed using the unitary operator $U_e^{-1}$ on the rung $1+r$ as a component of $\mathcal{U}_d^{(1+r,j)}$, without any change in the $V_d^{(1+r,j)}$ and  $R_d^{(1+r,j)}$ components of $\mathcal{U}_d^{(1+r,j)}$. Note, however, that the use of $U_e^{-1}$ would imply a change in the sign of $J_\perp$, which corresponds to a change of the type of spin-spin interactions on a rung from AFM to FM. We assume the more \emph{physical} scenario where the type of interaction between the spins is fixed throughout the protocol.  

 \item The protocol can also be designed with an encoding unitary $\mathcal{U}_e$ arising out of a rung Hamiltonian $\mathcal{H}_1(w)$ for an arbitrary value of the field-strength  $w$. The effect of this change  manifests in a change of the forms of  $R_d^{(1+r,j)}$, where  $R_d{(1+r,j)}$  depends on $w$ as 
 \begin{eqnarray}
           R_d^{(1+r,j)} &=&\left\{ \begin{array}{cc}
              \text{e}^{-\text{i} w\pi \sigma^z_{(1+r,j)} }, & j=1 \\
              \text{e}^{-\text{i} (w-\frac{1}{2})\pi \sigma^z_{(1+r,j)} }, & j=2
           \end{array}\right.
           \end{eqnarray}
upto a global phase. 
\end{enumerate}

The single-qubit TF, in this case, is computed as $f^\prime=\langle\psi_{1,1}|\rho_{1+r,j}|\psi_{1,1}\rangle$. It is logical to ask how $f^\prime$ is related to $f$ -- the TF of the effective 1D transfer of low-energy rung states, which is the central component of the above protocol.
To answer this, note that on the rung $(1+r)$, 
\begin{eqnarray}
\mathcal{U}_d\ket{\mathbf{0}}&=&\ket{0}^{\otimes L}, \\
\mathcal{U}_d\ket{\mathbf{1}}&=&\ket{1_j}\otimes_{j^\prime\neq j} \ket{0_{j^\prime}}, 
\end{eqnarray}
while the reduced density matrix of the receiver rung is of the generic form
\begin{eqnarray}\label{eq:rho_1+r}    \rho_{1+r}=\rho_0\ket{\mathbf{0}}\bra{\mathbf{0}}+\rho_{01}\ket{\mathbf{0}}\bra{\mathbf{1}}+\rho_{01}^*\ket{\mathbf{1}}\bra{\mathbf{0}}+\rho_{1}\ket{\mathbf{1}}\bra{\mathbf{1}}.  \nonumber \\
\end{eqnarray}
Therefore, 
\begin{eqnarray}\label{eq:rho_1+r,j}
\rho_{1+r,j}&=&\text{Tr}_{\overline{1+r,j}}(\mathcal{U}_d^{1+r}\rho_{1+r}{\mathcal{U}_d^{1+r}}^\dagger) \nonumber \\
&=&\rho_0\ket{0}\bra{0}+\rho_{01}\ket{0}\bra{1}+\rho_{01}^*\ket{1}\bra{0}+\rho_{1}\ket{1}\bra{1}, \nonumber \\
\end{eqnarray}
Thus for a given single qubit state $\ket{\psi_{i,j}}$ (see (\ref{eq:single_qubit_states}))  and its corresponding encoded low energy rung state $\ket{\psi_i}$ of the form~(\ref{eq:rung_initial_state}),  
\begin{eqnarray}
    \bra{\psi_{i,j}} \rho_{1+r,j} \ket{\psi_{i,j}}=\bra{\psi_i} \rho_{1+r} \ket{\psi_i}, 
\end{eqnarray}
i.e., $f^\prime=f$. Note that a deviation from the perturbation regime of the system parameters may induce an error in the fidelity approximated by the effective 1D XXZ transfer. However, as long as this error is small,  $f^\prime\approx f$. Note also that this result is independent of the value of $L$, and is valid for any protocol that involves encoding and decoding unitaries confined to only the input and output rungs, respectively.

For a fixed set of system parameters $w_c,\delta w, u$, and $v$ in the perturbation regime, the average TF at time $t$ for a specific transfer distance $r$ is computed as
\begin{eqnarray}
    \langle f^\prime\rangle &=& \int f^\prime P(f^\prime)df^\prime,   
\end{eqnarray}
where $P(f^\prime)$ is the probability distribution of $f^\prime$, obtained over a Haar uniformly generated sample of initial states $\ket{\psi_{1,1}}$. We plot the MATF, $\langle f^\prime\rangle_m$, for the single-qubit state transfer as a function of $r$ in the case of the two-leg ladder in Fig.~\ref{fig:single_qubit}, which shows a decreasing trend similar to the rung-to-rung transfer discussed in Sec.~\ref{sec:rung2rung_state_transfer}. Note that one can also attempt to transfer the input state~(\ref{eq:single_qubit_rung_state}) via a time evolution generated by the quasi-1D isotropic Heisenberg Hamiltonian $H(u,v,\delta w)$, where the system parameters $(u,v,\delta w)$ are small so that the perturbation regime remains valid. Therefore, a comparison between the MATF of the arbitrary single-qubit state transferred using the proposed protocol, and the MATF obtained via a transfer using the quasi-1D isotropic Heisenberg Hamiltonian is in order here. We present the data corresponding to this in Fig.~\ref{fig:single_qubit}, and note that 
\begin{enumerate} 
\item[(a)] MATF computed using the full ladder Hamiltonian is always less than the same obtained using the proposed protocol, thereby establishing the utility of the protocol, and 
\item[(b)] the proposed protocol is designed in such a way that the TF is independent of the position of the receiver qubit on the rung $1+r$, while in the case of the single-qubit state transfer using the full quasi-1D ladder Hamiltonian, it is not so. 
\end{enumerate}

\subsection{Single-qubit state transfer in four-leg ladder}

The non-trivial nature of the evolution generated by $\mathcal{H}_i(w_c)$  makes generalizing the encoding and decoding of the rung states difficult for increasing $L$. We further provide a specific protocol for single-qubit state transfer using a four-leg ladder, as follows. For ease of representation, we assume that Alice has the qubit $(1,1)$ in her possession. A schematic representation of the protocol is given in Fig.~\ref{fig:circuit}(b).

\vspace{0.25cm}
\hrule 
\vspace{0.25cm}

\noindent\textbf{Protocol for four-leg ladder}  

\begin{enumerate}
    \item[\textbf{1:}] \emph{Initialize.} Initialise Alice's qubit on the rung $1$ in the state $\ket{\psi_{1,1}}$ of the form~(\ref{eq:single_qubit_states}), and all other qubits  in the state $\ket{0}$, such that the state on rung $1$ is 
    \begin{eqnarray}
    \ket{\psi_1} &=& c_0\ket{0_{1,1}}\otimes_{j\neq 1}\ket{0_{1,j}}+c_1\ket{1_{1,1}}\otimes_{j\neq1}\ket{0_{1,j}}. 
    \end{eqnarray}
    The initial state of the system at $t=0$ is of the form~(\ref{eq:system_state_input}) with $\ket{\mathbf{0}_{i}}=\otimes_j\ket{0_{i,j}}$.  
    \item[\textbf{2:}] \emph{Encode.} The encoding of the initial state on the rung $1$ into the form~(\ref{eq:low_energy_rung_state_L2}) is done by an unitary operator with three components:   
    \begin{enumerate} 
           \item[\textbf{(a)}] Evolve $\ket{\psi_{1}}$ upto $t=\pi/2$ using the rung Hamiltonian $\mathcal{H}_1(w_c)$ (see Eq.~(\ref{eq:rung_hamiltonian})) such that $\ket{\psi_1}\rightarrow U_e^{(1)}\ket{\psi_1}$ with $U_e^{(1)}$ given in~(\ref{eq:rung_evolution}). The explicit form of the time-evolved rung state is 
           \begin{eqnarray}
              \ket{\psi_1} &=& c_0\ket{0000}+\frac{c_1 \text{e}^{-\text{i}3\pi/2} }{2}(\text{i}\ket{0001}+\text{i}\ket{0100}\nonumber\\
                 &&-\ket{1000}+\ket{0010}),
           \end{eqnarray}
           up to a global phase $\text{e}^{\text{i}3\pi/2}$. 

          \item[\textbf{(b)}] Apply $\sigma^z$ gate on the qubit $(1,1)$ such that $\ket{\psi_1}\rightarrow \sigma^z_{(1,1)}\ket{\psi_1}$.
           \item[\textbf{(c)}] Apply the single-qubit unitary operator $V_e^{(1,j)}$ (Eq.~(\ref{eq:encoding_unitary_single_qubit})) on the data qubits $(1,2)$ and $(1,4)$ to transform the state $\ket{\psi_1}\rightarrow V_e^{(1,2)}V_e^{(1,4)}\ket{\psi_1}$. The   encoded state, at this point, is
           \begin{eqnarray}
                 \ket{\psi_1} &=&  c_0\ket{0000}+\frac{c_1\text{e}^{-\text{i}\pi/2}}{2}(\ket{0001}+\ket{0100}\nonumber\\
                 &&-\ket{1000}-\ket{0010}), 
            \end{eqnarray} 
    \end{enumerate}       
    upto a global phase $\text{e}^{3\text{i}\pi/2}$, and the full encoding unitary $U_e^{(1)}$ on the rung $1$ is given by  
    \begin{eqnarray}
        \mathcal{U}_e^{(1)}=V_e^{(1,4)}V_e^{(1,2)}\sigma^z_{(1,1)}U_e^{(1)}.
    \end{eqnarray}

    \item[\textbf{3:}] \emph{Transfer.} The low-energy rung state $\ket{\psi_1}$ is now transferred to the rung $1+r$ using the unitary evolution $T$ (see Eq.~(\ref{eq:system_time_evolution})). See R--R transfer protocol in Sec.~\ref{sec:rung2rung_state_transfer} for details.

    \item[\textbf{4:}] \emph{Decode.} The decoding unitary $\mathcal{U}_d^{(1+r)}$ on the rung $1+r$ has the following components: 
    \begin{enumerate} 
           \item[\textbf{(a)}] Given that Bob wants to extract the single-qubit state from the qubit $(1+r,j)$, apply $\sigma^z$ on the qubit $(1+r,j)$, and the unitary operator $V_d^{(1+r,j^\prime)}$ (Eq.~(\ref{eq:decoding_unitary_single_qubit})) on the qubit $(1+r,j^\prime)$. If $j=1(3)$, $j^\prime=2$ and $4$, while for $j=2(4)$, $j^\prime=1$ and $3$.

           \item[\textbf{(b)}] Turn on the rung Hamiltonian $\mathcal{H}(w_c)$ on the rung  $1+r$ to evolve the state $\ket{\Psi}\rightarrow U_d^{(1+r)}\ket{\Psi}$, with $U_d^{(1+r)}$ given by (\ref{eq:decoding_rung_unitary}). 

           \item[\textbf{(c)}] Apply a local rotation $R_d^{(1+r,j)}$ on the data qubit $(1+r,j)$, such that $\ket{\Psi}\rightarrow R_d^{(1+r,j)}\ket{\Psi}$. The form of $R_d^{(1+r,j)}$ is given by  
           \begin{eqnarray}
           R_d^{(1+r,j)} &=&\left\{ \begin{array}{cc}
              I_{(1+r,j)}, & j=1 \text{ or } j=3 \\
              \sigma^z_{(1+r,j)} , & j=2 \text{ or }j=4
           \end{array}\right.
           \end{eqnarray}
    \end{enumerate} 
    The full decoding unitary is therefore given by 
    \begin{eqnarray}
        \mathcal{U}_d^{(1+r)}=R_d^{(1+r,j)}U_d^{(1+r)}V_d^{(1+r,j^\prime)}\sigma^{z}_{(1+r,j)}.
    \end{eqnarray}

    \item[\textbf{5:}] \emph{Extract.} Bob determines the state $\rho_{1+r,j}$ on the qubit $(1+r,j)$, $j=1,2$, as 
    \begin{eqnarray}
        \rho_{1+r,j}=\text{Tr}_{\overline{1+r,j}}\ket{\Psi}\bra{\Psi}
    \end{eqnarray}
    via tracing out all other qubits in the system except the qubit $(1+r,j)$. 
\end{enumerate}

\vspace{0.25cm}
\hrule 
\vspace{0.25cm}

\noindent As discussed in the case of the protocol for two-leg ladder, the TF $f^\prime= f$ for this protocol also. Consequently,  the variations of $\langle f^\prime\rangle_m$ as a function of $r$ is qualitatively similar to the same for the two-leg case. Also, similar to the protocol for the two-leg ladder, one can design the operators $R_d^{(1+r,j)}$ in terms of an arbitrary magnetic field $w$ on the qubits in the rung $(1+r)$ as 
           \begin{eqnarray}
           R_d^{(1+r,j)} &=&\left\{ \begin{array}{cc}
              \text{e}^{-\text{i} w\pi \sigma^z_{(1+r,j)} }, & j=1 \text{ or } j=3 \\
              \text{e}^{-\text{i} (w-\frac{1}{2})\pi \sigma^z_{(1+r,j)} }, & j=2 \text{ or }j=4
           \end{array}\right.
           \end{eqnarray}
We again emphasise here that determination of the encoding and the decoding unitary becomes difficult with increasing $L$, and due to the non-trivial nature of the time-evolution of a high-energy state generated by the rung Hamiltonian. However, as long as the designed $\mathcal{U}_e$ and $\mathcal{U}_d$ are confined respectively to the input and the receiver rungs, the TF for the single-qubit state transfer protocol can be represented by the TF of the transfer of an arbitrary single-qubit state from one lattice site to another using a 1D XXZ Hamiltonian.


\section{Conclusion and Outlook}
\label{sec:outlook}

In this paper, we explore the transfer of single- and multi-qubit quantum states on quasi-1D lattices, using only time evolutions via 1D Hamiltonians. For this, we employ the isotropic Heisenberg model in a magnetic field under the strong rung coupling limit, which can be mapped to a 1D XXZ model. We study the transfer of low energy states on a rung from one rung to another in the quasi-1D lattice in terms of the transfer of an arbitrary single-qubit state from one lattice site to another on the 1D lattice on which the effective 1D XXZ model is defined. Based on this, we propose specific encoding of single-qubit states into the low-energy states of rungs, and specific decoding of the transferred rung state to extract the single-qubit state,  such that the effective 1D transfer of the rung state via the 1D XXZ model can also be utilized for the arbitrary single-qubit state transfer. The proposed encoding and decoding involves time evolution of the initial state via the 1D rung Hamiltonian, and single-qubit phase gates on selected qubits in the sender and receiver rungs. We demonstrate that the transfer fidelity of the proposed protocol is equal to the transfer fidelity for an arbitrary state using 1D XXZ model, and is  always better than the same corresponding to the transfer of the single-qubit state via the \emph{bare} quasi-1D Heisenberg Hamiltonian with strong rung couplings.

We conclude with an outlook towards possible research directions originating from this work. While our protocol for single-qubit state transfer considers PBC along the rungs, it would be interesting to explore possible encoding and decoding of rung states with OBC along the rungs also, such that the effective 1D rung-to-rung low-energy state transfer can be utilized.  Moreover, the strong rung-coupling limit of the isotropic Heisenberg model can also be mapped to a 1D model by tuning the magnetic field at a special value such that each lattice site hosts a Hilbert space of dimension greater than two. Therefore, effective 1D transfer of states of higher dimensional objects, such as a qutrit, or a qudit, is also possible on a quasi-1D lattice, while the specific protocol remains to be worked out. It is also important to note that a strong-coupling expansion reduces the complexity, more specifically, the effective Hilbert space dimension in a number of lattice models~\cite{Mila2011}, and it would be interesting to develop state transfer protocols specific to these models.

\acknowledgements 

We acknowledge the use of \href{https://github.com/titaschanda/QIClib}{QIClib} -- a modern C++ library for general purpose quantum information processing and quantum computing. We also thank Dr. Prithvi Narayan P. and Prof. Aditi Sen (De) for useful discussions.

\twocolumngrid 

\bibliography{ref_njp}{}

\begin{thebibliography}{93}%
\makeatletter
\providecommand \@ifxundefined [1]{%
 \@ifx{#1\undefined}
}%
\providecommand \@ifnum [1]{%
 \ifnum #1\expandafter \@firstoftwo
 \else \expandafter \@secondoftwo
 \fi
}%
\providecommand \@ifx [1]{%
 \ifx #1\expandafter \@firstoftwo
 \else \expandafter \@secondoftwo
 \fi
}%
\providecommand \natexlab [1]{#1}%
\providecommand \enquote  [1]{``#1''}%
\providecommand \bibnamefont  [1]{#1}%
\providecommand \bibfnamefont [1]{#1}%
\providecommand \citenamefont [1]{#1}%
\providecommand \href@noop [0]{\@secondoftwo}%
\providecommand \href [0]{\begingroup \@sanitize@url \@href}%
\providecommand \@href[1]{\@@startlink{#1}\@@href}%
\providecommand \@@href[1]{\endgroup#1\@@endlink}%
\providecommand \@sanitize@url [0]{\catcode `\\12\catcode `\$12\catcode
  `\&12\catcode `\#12\catcode `\^12\catcode `\_12\catcode `\%12\relax}%
\providecommand \@@startlink[1]{}%
\providecommand \@@endlink[0]{}%
\providecommand \url  [0]{\begingroup\@sanitize@url \@url }%
\providecommand \@url [1]{\endgroup\@href {#1}{\urlprefix }}%
\providecommand \urlprefix  [0]{URL }%
\providecommand \Eprint [0]{\href }%
\providecommand \doibase [0]{http://dx.doi.org/}%
\providecommand \selectlanguage [0]{\@gobble}%
\providecommand \bibinfo  [0]{\@secondoftwo}%
\providecommand \bibfield  [0]{\@secondoftwo}%
\providecommand \translation [1]{[#1]}%
\providecommand \BibitemOpen [0]{}%
\providecommand \bibitemStop [0]{}%
\providecommand \bibitemNoStop [0]{.\EOS\space}%
\providecommand \EOS [0]{\spacefactor3000\relax}%
\providecommand \BibitemShut  [1]{\csname bibitem#1\endcsname}%
\let\auto@bib@innerbib\@empty
\bibitem [{\citenamefont {Nielsen}\ and\ \citenamefont
  {Chuang}(2010)}]{nielsen2010}%
  \BibitemOpen
  \bibfield  {author} {\bibinfo {author} {\bibfnamefont {M.~A.}\ \bibnamefont
  {Nielsen}}\ and\ \bibinfo {author} {\bibfnamefont {I.~L.}\ \bibnamefont
  {Chuang}},\ }\href@noop {} {\emph {\bibinfo {title} {Quantum Computation and
  Quantum Information}}}\ (\bibinfo  {publisher} {Cambridge University Press},\
  \bibinfo {year} {2010})\BibitemShut {NoStop}%
\bibitem [{\citenamefont {Wilde}(2017)}]{wilde_book}%
  \BibitemOpen
  \bibfield  {author} {\bibinfo {author} {\bibfnamefont {M.~M.}\ \bibnamefont
  {Wilde}},\ }\href@noop {} {\emph {\bibinfo {title} {Quantum Information
  Theory, 2nd ed.}}}\ (\bibinfo  {publisher} {Cambridge University Press},\
  \bibinfo {address} {Cambridge, UK},\ \bibinfo {year} {2017})\BibitemShut
  {NoStop}%
\bibitem [{\citenamefont {Amico}\ \emph {et~al.}(2008)\citenamefont {Amico},
  \citenamefont {Fazio}, \citenamefont {Osterloh},\ and\ \citenamefont
  {Vedral}}]{Amico2008}%
  \BibitemOpen
  \bibfield  {author} {\bibinfo {author} {\bibfnamefont {L.}~\bibnamefont
  {Amico}}, \bibinfo {author} {\bibfnamefont {R.}~\bibnamefont {Fazio}},
  \bibinfo {author} {\bibfnamefont {A.}~\bibnamefont {Osterloh}}, \ and\
  \bibinfo {author} {\bibfnamefont {V.}~\bibnamefont {Vedral}},\ }\href
  {\doibase 10.1103/RevModPhys.80.517} {\bibfield  {journal} {\bibinfo
  {journal} {Rev. Mod. Phys.}\ }\textbf {\bibinfo {volume} {80}},\ \bibinfo
  {pages} {517} (\bibinfo {year} {2008})}\BibitemShut {NoStop}%
\bibitem [{\citenamefont {Latorre}\ and\ \citenamefont
  {Riera}(2009)}]{Latorre_2009}%
  \BibitemOpen
  \bibfield  {author} {\bibinfo {author} {\bibfnamefont {J.~I.}\ \bibnamefont
  {Latorre}}\ and\ \bibinfo {author} {\bibfnamefont {A.}~\bibnamefont
  {Riera}},\ }\href {\doibase 10.1088/1751-8113/42/50/504002} {\bibfield
  {journal} {\bibinfo  {journal} {Journal of Physics A: Mathematical and
  Theoretical}\ }\textbf {\bibinfo {volume} {42}},\ \bibinfo {pages} {504002}
  (\bibinfo {year} {2009})}\BibitemShut {NoStop}%
\bibitem [{\citenamefont {Modi}\ \emph {et~al.}(2012)\citenamefont {Modi},
  \citenamefont {Brodutch}, \citenamefont {Cable}, \citenamefont {Paterek},\
  and\ \citenamefont {Vedral}}]{Modi2012}%
  \BibitemOpen
  \bibfield  {author} {\bibinfo {author} {\bibfnamefont {K.}~\bibnamefont
  {Modi}}, \bibinfo {author} {\bibfnamefont {A.}~\bibnamefont {Brodutch}},
  \bibinfo {author} {\bibfnamefont {H.}~\bibnamefont {Cable}}, \bibinfo
  {author} {\bibfnamefont {T.}~\bibnamefont {Paterek}}, \ and\ \bibinfo
  {author} {\bibfnamefont {V.}~\bibnamefont {Vedral}},\ }\href {\doibase
  10.1103/RevModPhys.84.1655} {\bibfield  {journal} {\bibinfo  {journal} {Rev.
  Mod. Phys.}\ }\textbf {\bibinfo {volume} {84}},\ \bibinfo {pages} {1655}
  (\bibinfo {year} {2012})}\BibitemShut {NoStop}%
\bibitem [{\citenamefont {Laflorencie}(2016)}]{laflorencie2016}%
  \BibitemOpen
  \bibfield  {author} {\bibinfo {author} {\bibfnamefont {N.}~\bibnamefont
  {Laflorencie}},\ }\href {\doibase
  https://doi.org/10.1016/j.physrep.2016.06.008} {\bibfield  {journal}
  {\bibinfo  {journal} {Physics Reports}\ }\textbf {\bibinfo {volume} {646}},\
  \bibinfo {pages} {1} (\bibinfo {year} {2016})},\ \bibinfo {note} {quantum
  entanglement in condensed matter systems}\BibitemShut {NoStop}%
\bibitem [{\citenamefont {Chiara}\ and\ \citenamefont
  {Sanpera}(2018)}]{DeChiara_2018}%
  \BibitemOpen
  \bibfield  {author} {\bibinfo {author} {\bibfnamefont {G.~D.}\ \bibnamefont
  {Chiara}}\ and\ \bibinfo {author} {\bibfnamefont {A.}~\bibnamefont
  {Sanpera}},\ }\href {\doibase 10.1088/1361-6633/aabf61} {\bibfield  {journal}
  {\bibinfo  {journal} {Reports on Progress in Physics}\ }\textbf {\bibinfo
  {volume} {81}},\ \bibinfo {pages} {074002} (\bibinfo {year}
  {2018})}\BibitemShut {NoStop}%
\bibitem [{\citenamefont {Bera}\ \emph {et~al.}(2017)\citenamefont {Bera},
  \citenamefont {Das}, \citenamefont {Sadhukhan}, \citenamefont {Roy},
  \citenamefont {Sen(De)},\ and\ \citenamefont {Sen}}]{Bera_2018}%
  \BibitemOpen
  \bibfield  {author} {\bibinfo {author} {\bibfnamefont {A.}~\bibnamefont
  {Bera}}, \bibinfo {author} {\bibfnamefont {T.}~\bibnamefont {Das}}, \bibinfo
  {author} {\bibfnamefont {D.}~\bibnamefont {Sadhukhan}}, \bibinfo {author}
  {\bibfnamefont {S.~S.}\ \bibnamefont {Roy}}, \bibinfo {author} {\bibfnamefont
  {A.}~\bibnamefont {Sen(De)}}, \ and\ \bibinfo {author} {\bibfnamefont
  {U.}~\bibnamefont {Sen}},\ }\href {\doibase 10.1088/1361-6633/aa872f}
  {\bibfield  {journal} {\bibinfo  {journal} {Reports on Progress in Physics}\
  }\textbf {\bibinfo {volume} {81}},\ \bibinfo {pages} {024001} (\bibinfo
  {year} {2017})}\BibitemShut {NoStop}%
\bibitem [{\citenamefont {Bose}(2003)}]{Bose2003}%
  \BibitemOpen
  \bibfield  {author} {\bibinfo {author} {\bibfnamefont {S.}~\bibnamefont
  {Bose}},\ }\href {\doibase 10.1103/PhysRevLett.91.207901} {\bibfield
  {journal} {\bibinfo  {journal} {Phys. Rev. Lett.}\ }\textbf {\bibinfo
  {volume} {91}},\ \bibinfo {pages} {207901} (\bibinfo {year}
  {2003})}\BibitemShut {NoStop}%
\bibitem [{\citenamefont {Bose}\ \emph {et~al.}(2013)\citenamefont {Bose},
  \citenamefont {Bayat}, \citenamefont {Sodano}, \citenamefont {Banchi},\ and\
  \citenamefont {Verrucchi}}]{Bose2013_chapter}%
  \BibitemOpen
  \bibfield  {author} {\bibinfo {author} {\bibfnamefont {S.}~\bibnamefont
  {Bose}}, \bibinfo {author} {\bibfnamefont {A.}~\bibnamefont {Bayat}},
  \bibinfo {author} {\bibfnamefont {P.}~\bibnamefont {Sodano}}, \bibinfo
  {author} {\bibfnamefont {L.}~\bibnamefont {Banchi}}, \ and\ \bibinfo {author}
  {\bibfnamefont {P.}~\bibnamefont {Verrucchi}},\ }\enquote {\bibinfo {title}
  {Spin chains as data buses, logic buses and entanglers},}\ in\ \href
  {\doibase https://doi.org/10.1007/978-3-642-39937-4} {\emph {\bibinfo
  {booktitle} {Quantum State Transfer and Network Engineering}}},\ \bibinfo
  {editor} {edited by\ \bibinfo {editor} {\bibfnamefont {G.~M.}\ \bibnamefont
  {Nikolopoulos}}\ and\ \bibinfo {editor} {\bibfnamefont {I.}~\bibnamefont
  {Jex}}}\ (\bibinfo  {publisher} {Springer Berlin, Heidelberg},\ \bibinfo
  {year} {2013})\ Chap.~\bibinfo {chapter} {1}, pp.\ \bibinfo {pages}
  {1--38}\BibitemShut {NoStop}%
\bibitem [{\citenamefont {Raussendorf}\ and\ \citenamefont
  {Briegel}(2001)}]{raussendorf2001}%
  \BibitemOpen
  \bibfield  {author} {\bibinfo {author} {\bibfnamefont {R.}~\bibnamefont
  {Raussendorf}}\ and\ \bibinfo {author} {\bibfnamefont {H.~J.}\ \bibnamefont
  {Briegel}},\ }\href {\doibase 10.1103/PhysRevLett.86.5188} {\bibfield
  {journal} {\bibinfo  {journal} {Phys. Rev. Lett.}\ }\textbf {\bibinfo
  {volume} {86}},\ \bibinfo {pages} {5188} (\bibinfo {year}
  {2001})}\BibitemShut {NoStop}%
\bibitem [{\citenamefont {Raussendorf}\ \emph {et~al.}(2003)\citenamefont
  {Raussendorf}, \citenamefont {Browne},\ and\ \citenamefont
  {Briegel}}]{raussendorf2003}%
  \BibitemOpen
  \bibfield  {author} {\bibinfo {author} {\bibfnamefont {R.}~\bibnamefont
  {Raussendorf}}, \bibinfo {author} {\bibfnamefont {D.~E.}\ \bibnamefont
  {Browne}}, \ and\ \bibinfo {author} {\bibfnamefont {H.~J.}\ \bibnamefont
  {Briegel}},\ }\href {\doibase 10.1103/PhysRevA.68.022312} {\bibfield
  {journal} {\bibinfo  {journal} {Phys. Rev. A}\ }\textbf {\bibinfo {volume}
  {68}},\ \bibinfo {pages} {022312} (\bibinfo {year} {2003})}\BibitemShut
  {NoStop}%
\bibitem [{\citenamefont {Briegel}\ \emph {et~al.}(2009)\citenamefont
  {Briegel}, \citenamefont {Browne}, \citenamefont {D{\"u}r}, \citenamefont
  {Raussendorf},\ and\ \citenamefont {Van~den Nest}}]{briegel2009}%
  \BibitemOpen
  \bibfield  {author} {\bibinfo {author} {\bibfnamefont {H.~J.}\ \bibnamefont
  {Briegel}}, \bibinfo {author} {\bibfnamefont {D.~E.}\ \bibnamefont {Browne}},
  \bibinfo {author} {\bibfnamefont {W.}~\bibnamefont {D{\"u}r}}, \bibinfo
  {author} {\bibfnamefont {R.}~\bibnamefont {Raussendorf}}, \ and\ \bibinfo
  {author} {\bibfnamefont {M.}~\bibnamefont {Van~den Nest}},\ }\href {\doibase
  10.1038/nphys1157} {\bibfield  {journal} {\bibinfo  {journal} {Nat. Phys.}\
  }\textbf {\bibinfo {volume} {5}},\ \bibinfo {pages} {19} (\bibinfo {year}
  {2009})}\BibitemShut {NoStop}%
\bibitem [{\citenamefont {Wei}(2018)}]{Wei2018}%
  \BibitemOpen
  \bibfield  {author} {\bibinfo {author} {\bibfnamefont {T.-C.}\ \bibnamefont
  {Wei}},\ }\href {\doibase 10.1080/23746149.2018.1461026} {\bibfield
  {journal} {\bibinfo  {journal} {Advances in Physics: X}\ }\textbf {\bibinfo
  {volume} {3}},\ \bibinfo {pages} {1461026} (\bibinfo {year}
  {2018})}\BibitemShut {NoStop}%
\bibitem [{\citenamefont {Dennis}\ \emph {et~al.}(2002)\citenamefont {Dennis},
  \citenamefont {Kitaev}, \citenamefont {Landahl},\ and\ \citenamefont
  {Preskill}}]{kitaev2001}%
  \BibitemOpen
  \bibfield  {author} {\bibinfo {author} {\bibfnamefont {E.}~\bibnamefont
  {Dennis}}, \bibinfo {author} {\bibfnamefont {A.}~\bibnamefont {Kitaev}},
  \bibinfo {author} {\bibfnamefont {A.}~\bibnamefont {Landahl}}, \ and\
  \bibinfo {author} {\bibfnamefont {J.}~\bibnamefont {Preskill}},\ }\href
  {\doibase 10.1063/1.1499754} {\bibfield  {journal} {\bibinfo  {journal} {J.
  Math. Phys.}\ }\textbf {\bibinfo {volume} {43}},\ \bibinfo {pages} {4452}
  (\bibinfo {year} {2002})}\BibitemShut {NoStop}%
\bibitem [{\citenamefont {Kitaev}(2006)}]{kitaev2006}%
  \BibitemOpen
  \bibfield  {author} {\bibinfo {author} {\bibfnamefont {A.}~\bibnamefont
  {Kitaev}},\ }\href {\doibase https://doi.org/10.1016/j.aop.2005.10.005}
  {\bibfield  {journal} {\bibinfo  {journal} {Ann. Phys.}\ }\textbf {\bibinfo
  {volume} {321}},\ \bibinfo {pages} {2 } (\bibinfo {year} {2006})}\BibitemShut
  {NoStop}%
\bibitem [{\citenamefont {Bombin}\ and\ \citenamefont
  {Martin-Delgado}(2006)}]{bombin2006}%
  \BibitemOpen
  \bibfield  {author} {\bibinfo {author} {\bibfnamefont {H.}~\bibnamefont
  {Bombin}}\ and\ \bibinfo {author} {\bibfnamefont {M.~A.}\ \bibnamefont
  {Martin-Delgado}},\ }\href {\doibase 10.1103/PhysRevLett.97.180501}
  {\bibfield  {journal} {\bibinfo  {journal} {Phys. Rev. Lett.}\ }\textbf
  {\bibinfo {volume} {97}},\ \bibinfo {pages} {180501} (\bibinfo {year}
  {2006})}\BibitemShut {NoStop}%
\bibitem [{\citenamefont {Bombin}\ and\ \citenamefont
  {Martin-Delgado}(2007)}]{bombin2007}%
  \BibitemOpen
  \bibfield  {author} {\bibinfo {author} {\bibfnamefont {H.}~\bibnamefont
  {Bombin}}\ and\ \bibinfo {author} {\bibfnamefont {M.~A.}\ \bibnamefont
  {Martin-Delgado}},\ }\href {\doibase 10.1103/PhysRevLett.98.160502}
  {\bibfield  {journal} {\bibinfo  {journal} {Phys. Rev. Lett.}\ }\textbf
  {\bibinfo {volume} {98}},\ \bibinfo {pages} {160502} (\bibinfo {year}
  {2007})}\BibitemShut {NoStop}%
\bibitem [{\citenamefont {Porras}\ and\ \citenamefont
  {Cirac}(2004)}]{Porras2004}%
  \BibitemOpen
  \bibfield  {author} {\bibinfo {author} {\bibfnamefont {D.}~\bibnamefont
  {Porras}}\ and\ \bibinfo {author} {\bibfnamefont {J.~I.}\ \bibnamefont
  {Cirac}},\ }\href {\doibase 10.1103/PhysRevLett.92.207901} {\bibfield
  {journal} {\bibinfo  {journal} {Phys. Rev. Lett.}\ }\textbf {\bibinfo
  {volume} {92}},\ \bibinfo {pages} {207901} (\bibinfo {year}
  {2004})}\BibitemShut {NoStop}%
\bibitem [{\citenamefont {Leibfried}\ \emph {et~al.}(2005)\citenamefont
  {Leibfried}, \citenamefont {Knill}, \citenamefont {Seidelin}, \citenamefont
  {Britton}, \citenamefont {Blakestad}, \citenamefont {Chiaverini},
  \citenamefont {Hume}, \citenamefont {Itano}, \citenamefont {Jost},
  \citenamefont {Langer}, \citenamefont {Ozeri}, \citenamefont {Reichle},\ and\
  \citenamefont {Wineland}}]{Leibfried2005}%
  \BibitemOpen
  \bibfield  {author} {\bibinfo {author} {\bibfnamefont {D.}~\bibnamefont
  {Leibfried}}, \bibinfo {author} {\bibfnamefont {E.}~\bibnamefont {Knill}},
  \bibinfo {author} {\bibfnamefont {S.}~\bibnamefont {Seidelin}}, \bibinfo
  {author} {\bibfnamefont {J.}~\bibnamefont {Britton}}, \bibinfo {author}
  {\bibfnamefont {R.~B.}\ \bibnamefont {Blakestad}}, \bibinfo {author}
  {\bibfnamefont {J.}~\bibnamefont {Chiaverini}}, \bibinfo {author}
  {\bibfnamefont {D.~B.}\ \bibnamefont {Hume}}, \bibinfo {author}
  {\bibfnamefont {W.~M.}\ \bibnamefont {Itano}}, \bibinfo {author}
  {\bibfnamefont {J.~D.}\ \bibnamefont {Jost}}, \bibinfo {author}
  {\bibfnamefont {C.}~\bibnamefont {Langer}}, \bibinfo {author} {\bibfnamefont
  {R.}~\bibnamefont {Ozeri}}, \bibinfo {author} {\bibfnamefont
  {R.}~\bibnamefont {Reichle}}, \ and\ \bibinfo {author} {\bibfnamefont
  {D.~J.}\ \bibnamefont {Wineland}},\ }\href {\doibase 10.1038/nature04251}
  {\bibfield  {journal} {\bibinfo  {journal} {Nature}\ }\textbf {\bibinfo
  {volume} {438}},\ \bibinfo {pages} {639} (\bibinfo {year}
  {2005})}\BibitemShut {NoStop}%
\bibitem [{\citenamefont {Monz}\ \emph {et~al.}(2011)\citenamefont {Monz},
  \citenamefont {Schindler}, \citenamefont {Barreiro}, \citenamefont {Chwalla},
  \citenamefont {Nigg}, \citenamefont {Coish}, \citenamefont {Harlander},
  \citenamefont {H\"ansel}, \citenamefont {Hennrich},\ and\ \citenamefont
  {Blatt}}]{monz2011}%
  \BibitemOpen
  \bibfield  {author} {\bibinfo {author} {\bibfnamefont {T.}~\bibnamefont
  {Monz}}, \bibinfo {author} {\bibfnamefont {P.}~\bibnamefont {Schindler}},
  \bibinfo {author} {\bibfnamefont {J.~T.}\ \bibnamefont {Barreiro}}, \bibinfo
  {author} {\bibfnamefont {M.}~\bibnamefont {Chwalla}}, \bibinfo {author}
  {\bibfnamefont {D.}~\bibnamefont {Nigg}}, \bibinfo {author} {\bibfnamefont
  {W.~A.}\ \bibnamefont {Coish}}, \bibinfo {author} {\bibfnamefont
  {M.}~\bibnamefont {Harlander}}, \bibinfo {author} {\bibfnamefont
  {W.}~\bibnamefont {H\"ansel}}, \bibinfo {author} {\bibfnamefont
  {M.}~\bibnamefont {Hennrich}}, \ and\ \bibinfo {author} {\bibfnamefont
  {R.}~\bibnamefont {Blatt}},\ }\href {\doibase 10.1103/PhysRevLett.106.130506}
  {\bibfield  {journal} {\bibinfo  {journal} {Phys. Rev. Lett.}\ }\textbf
  {\bibinfo {volume} {106}},\ \bibinfo {pages} {130506} (\bibinfo {year}
  {2011})}\BibitemShut {NoStop}%
\bibitem [{\citenamefont {Korenblit}\ \emph {et~al.}(2012)\citenamefont
  {Korenblit}, \citenamefont {Kafri}, \citenamefont {Campbell}, \citenamefont
  {Islam}, \citenamefont {Edwards}, \citenamefont {Gong}, \citenamefont {Lin},
  \citenamefont {Duan}, \citenamefont {Kim}, \citenamefont {Kim},\ and\
  \citenamefont {Monroe}}]{Korenblit_2012}%
  \BibitemOpen
  \bibfield  {author} {\bibinfo {author} {\bibfnamefont {S.}~\bibnamefont
  {Korenblit}}, \bibinfo {author} {\bibfnamefont {D.}~\bibnamefont {Kafri}},
  \bibinfo {author} {\bibfnamefont {W.~C.}\ \bibnamefont {Campbell}}, \bibinfo
  {author} {\bibfnamefont {R.}~\bibnamefont {Islam}}, \bibinfo {author}
  {\bibfnamefont {E.~E.}\ \bibnamefont {Edwards}}, \bibinfo {author}
  {\bibfnamefont {Z.-X.}\ \bibnamefont {Gong}}, \bibinfo {author}
  {\bibfnamefont {G.-D.}\ \bibnamefont {Lin}}, \bibinfo {author} {\bibfnamefont
  {L.-M.}\ \bibnamefont {Duan}}, \bibinfo {author} {\bibfnamefont
  {J.}~\bibnamefont {Kim}}, \bibinfo {author} {\bibfnamefont {K.}~\bibnamefont
  {Kim}}, \ and\ \bibinfo {author} {\bibfnamefont {C.}~\bibnamefont {Monroe}},\
  }\href {\doibase 10.1088/1367-2630/14/9/095024} {\bibfield  {journal}
  {\bibinfo  {journal} {New Journal of Physics}\ }\textbf {\bibinfo {volume}
  {14}},\ \bibinfo {pages} {095024} (\bibinfo {year} {2012})}\BibitemShut
  {NoStop}%
\bibitem [{\citenamefont {Bohnet}\ \emph {et~al.}(2016)\citenamefont {Bohnet},
  \citenamefont {Sawyer}, \citenamefont {Britton}, \citenamefont {Wall},
  \citenamefont {Rey}, \citenamefont {Foss-Feig},\ and\ \citenamefont
  {Bollinger}}]{Bohnet2016}%
  \BibitemOpen
  \bibfield  {author} {\bibinfo {author} {\bibfnamefont {J.~G.}\ \bibnamefont
  {Bohnet}}, \bibinfo {author} {\bibfnamefont {B.~C.}\ \bibnamefont {Sawyer}},
  \bibinfo {author} {\bibfnamefont {J.~W.}\ \bibnamefont {Britton}}, \bibinfo
  {author} {\bibfnamefont {M.~L.}\ \bibnamefont {Wall}}, \bibinfo {author}
  {\bibfnamefont {A.~M.}\ \bibnamefont {Rey}}, \bibinfo {author} {\bibfnamefont
  {M.}~\bibnamefont {Foss-Feig}}, \ and\ \bibinfo {author} {\bibfnamefont
  {J.~J.}\ \bibnamefont {Bollinger}},\ }\href {\doibase
  10.1126/science.aad9958} {\bibfield  {journal} {\bibinfo  {journal}
  {Science}\ }\textbf {\bibinfo {volume} {352}},\ \bibinfo {pages} {1297}
  (\bibinfo {year} {2016})}\BibitemShut {NoStop}%
\bibitem [{\citenamefont {Barends}\ \emph {et~al.}(2014)\citenamefont
  {Barends}, \citenamefont {Kelly}, \citenamefont {Megrant}, \citenamefont
  {Veitia}, \citenamefont {Sank}, \citenamefont {Jeffrey}, \citenamefont
  {White}, \citenamefont {Mutus}, \citenamefont {Fowler}, \citenamefont
  {Campbell}, \citenamefont {Chen}, \citenamefont {Chen}, \citenamefont
  {Chiaro}, \citenamefont {Dunsworth}, \citenamefont {Neill}, \citenamefont
  {O'Malley}, \citenamefont {Roushan}, \citenamefont {Vainsencher},
  \citenamefont {Wenner}, \citenamefont {Korotkov}, \citenamefont {Cleland},\
  and\ \citenamefont {Martinis}}]{barends2014}%
  \BibitemOpen
  \bibfield  {author} {\bibinfo {author} {\bibfnamefont {R.}~\bibnamefont
  {Barends}}, \bibinfo {author} {\bibfnamefont {J.}~\bibnamefont {Kelly}},
  \bibinfo {author} {\bibfnamefont {A.}~\bibnamefont {Megrant}}, \bibinfo
  {author} {\bibfnamefont {A.}~\bibnamefont {Veitia}}, \bibinfo {author}
  {\bibfnamefont {D.}~\bibnamefont {Sank}}, \bibinfo {author} {\bibfnamefont
  {E.}~\bibnamefont {Jeffrey}}, \bibinfo {author} {\bibfnamefont {T.~C.}\
  \bibnamefont {White}}, \bibinfo {author} {\bibfnamefont {J.}~\bibnamefont
  {Mutus}}, \bibinfo {author} {\bibfnamefont {A.~G.}\ \bibnamefont {Fowler}},
  \bibinfo {author} {\bibfnamefont {B.}~\bibnamefont {Campbell}}, \bibinfo
  {author} {\bibfnamefont {Y.}~\bibnamefont {Chen}}, \bibinfo {author}
  {\bibfnamefont {Z.}~\bibnamefont {Chen}}, \bibinfo {author} {\bibfnamefont
  {B.}~\bibnamefont {Chiaro}}, \bibinfo {author} {\bibfnamefont
  {A.}~\bibnamefont {Dunsworth}}, \bibinfo {author} {\bibfnamefont
  {C.}~\bibnamefont {Neill}}, \bibinfo {author} {\bibfnamefont
  {P.}~\bibnamefont {O'Malley}}, \bibinfo {author} {\bibfnamefont
  {P.}~\bibnamefont {Roushan}}, \bibinfo {author} {\bibfnamefont
  {A.}~\bibnamefont {Vainsencher}}, \bibinfo {author} {\bibfnamefont
  {J.}~\bibnamefont {Wenner}}, \bibinfo {author} {\bibfnamefont {A.~N.}\
  \bibnamefont {Korotkov}}, \bibinfo {author} {\bibfnamefont {A.~N.}\
  \bibnamefont {Cleland}}, \ and\ \bibinfo {author} {\bibfnamefont {J.~M.}\
  \bibnamefont {Martinis}},\ }\href {\doibase 10.1038/nature13171} {\bibfield
  {journal} {\bibinfo  {journal} {Nature}\ }\textbf {\bibinfo {volume} {508}},\
  \bibinfo {pages} {500} (\bibinfo {year} {2014})}\BibitemShut {NoStop}%
\bibitem [{\citenamefont {Yanay}\ \emph {et~al.}(2020)\citenamefont {Yanay},
  \citenamefont {Braum{\"u}ller}, \citenamefont {Gustavsson}, \citenamefont
  {Oliver},\ and\ \citenamefont {Tahan}}]{Yariv2020}%
  \BibitemOpen
  \bibfield  {author} {\bibinfo {author} {\bibfnamefont {Y.}~\bibnamefont
  {Yanay}}, \bibinfo {author} {\bibfnamefont {J.}~\bibnamefont
  {Braum{\"u}ller}}, \bibinfo {author} {\bibfnamefont {S.}~\bibnamefont
  {Gustavsson}}, \bibinfo {author} {\bibfnamefont {W.~D.}\ \bibnamefont
  {Oliver}}, \ and\ \bibinfo {author} {\bibfnamefont {C.}~\bibnamefont
  {Tahan}},\ }\href {\doibase 10.1038/s41534-020-0269-1} {\bibfield  {journal}
  {\bibinfo  {journal} {npj Quantum Information}\ }\textbf {\bibinfo {volume}
  {6}},\ \bibinfo {pages} {58} (\bibinfo {year} {2020})}\BibitemShut {NoStop}%
\bibitem [{\citenamefont {Vandersypen}\ and\ \citenamefont
  {Chuang}(2005)}]{Vandersypen2005}%
  \BibitemOpen
  \bibfield  {author} {\bibinfo {author} {\bibfnamefont {L.~M.~K.}\
  \bibnamefont {Vandersypen}}\ and\ \bibinfo {author} {\bibfnamefont {I.~L.}\
  \bibnamefont {Chuang}},\ }\href {\doibase 10.1103/RevModPhys.76.1037}
  {\bibfield  {journal} {\bibinfo  {journal} {Rev. Mod. Phys.}\ }\textbf
  {\bibinfo {volume} {76}},\ \bibinfo {pages} {1037} (\bibinfo {year}
  {2005})}\BibitemShut {NoStop}%
\bibitem [{\citenamefont {Negrevergne}\ \emph {et~al.}(2006)\citenamefont
  {Negrevergne}, \citenamefont {Mahesh}, \citenamefont {Ryan}, \citenamefont
  {Ditty}, \citenamefont {Cyr-Racine}, \citenamefont {Power}, \citenamefont
  {Boulant}, \citenamefont {Havel}, \citenamefont {Cory},\ and\ \citenamefont
  {Laflamme}}]{Negrevergne2006}%
  \BibitemOpen
  \bibfield  {author} {\bibinfo {author} {\bibfnamefont {C.}~\bibnamefont
  {Negrevergne}}, \bibinfo {author} {\bibfnamefont {T.~S.}\ \bibnamefont
  {Mahesh}}, \bibinfo {author} {\bibfnamefont {C.~A.}\ \bibnamefont {Ryan}},
  \bibinfo {author} {\bibfnamefont {M.}~\bibnamefont {Ditty}}, \bibinfo
  {author} {\bibfnamefont {F.}~\bibnamefont {Cyr-Racine}}, \bibinfo {author}
  {\bibfnamefont {W.}~\bibnamefont {Power}}, \bibinfo {author} {\bibfnamefont
  {N.}~\bibnamefont {Boulant}}, \bibinfo {author} {\bibfnamefont
  {T.}~\bibnamefont {Havel}}, \bibinfo {author} {\bibfnamefont {D.~G.}\
  \bibnamefont {Cory}}, \ and\ \bibinfo {author} {\bibfnamefont
  {R.}~\bibnamefont {Laflamme}},\ }\href {\doibase
  10.1103/PhysRevLett.96.170501} {\bibfield  {journal} {\bibinfo  {journal}
  {Phys. Rev. Lett.}\ }\textbf {\bibinfo {volume} {96}},\ \bibinfo {pages}
  {170501} (\bibinfo {year} {2006})}\BibitemShut {NoStop}%
\bibitem [{\citenamefont {Schechter}\ and\ \citenamefont
  {Stamp}(2008)}]{Schechter2008}%
  \BibitemOpen
  \bibfield  {author} {\bibinfo {author} {\bibfnamefont {M.}~\bibnamefont
  {Schechter}}\ and\ \bibinfo {author} {\bibfnamefont {P.~C.~E.}\ \bibnamefont
  {Stamp}},\ }\href {\doibase 10.1103/PhysRevB.78.054438} {\bibfield  {journal}
  {\bibinfo  {journal} {Phys. Rev. B}\ }\textbf {\bibinfo {volume} {78}},\
  \bibinfo {pages} {054438} (\bibinfo {year} {2008})}\BibitemShut {NoStop}%
\bibitem [{\citenamefont {Bradley}\ \emph {et~al.}(2019)\citenamefont
  {Bradley}, \citenamefont {Randall}, \citenamefont {Abobeih}, \citenamefont
  {Berrevoets}, \citenamefont {Degen}, \citenamefont {Bakker}, \citenamefont
  {Markham}, \citenamefont {Twitchen},\ and\ \citenamefont
  {Taminiau}}]{Bradley2019}%
  \BibitemOpen
  \bibfield  {author} {\bibinfo {author} {\bibfnamefont {C.~E.}\ \bibnamefont
  {Bradley}}, \bibinfo {author} {\bibfnamefont {J.}~\bibnamefont {Randall}},
  \bibinfo {author} {\bibfnamefont {M.~H.}\ \bibnamefont {Abobeih}}, \bibinfo
  {author} {\bibfnamefont {R.~C.}\ \bibnamefont {Berrevoets}}, \bibinfo
  {author} {\bibfnamefont {M.~J.}\ \bibnamefont {Degen}}, \bibinfo {author}
  {\bibfnamefont {M.~A.}\ \bibnamefont {Bakker}}, \bibinfo {author}
  {\bibfnamefont {M.}~\bibnamefont {Markham}}, \bibinfo {author} {\bibfnamefont
  {D.~J.}\ \bibnamefont {Twitchen}}, \ and\ \bibinfo {author} {\bibfnamefont
  {T.~H.}\ \bibnamefont {Taminiau}},\ }\href {\doibase
  10.1103/PhysRevX.9.031045} {\bibfield  {journal} {\bibinfo  {journal} {Phys.
  Rev. X}\ }\textbf {\bibinfo {volume} {9}},\ \bibinfo {pages} {031045}
  (\bibinfo {year} {2019})}\BibitemShut {NoStop}%
\bibitem [{\citenamefont {Greiner}\ \emph {et~al.}(2002)\citenamefont
  {Greiner}, \citenamefont {Mandel}, \citenamefont {Esslinger}, \citenamefont
  {H{\"a}nsch},\ and\ \citenamefont {Bloch}}]{Greiner2002}%
  \BibitemOpen
  \bibfield  {author} {\bibinfo {author} {\bibfnamefont {M.}~\bibnamefont
  {Greiner}}, \bibinfo {author} {\bibfnamefont {O.}~\bibnamefont {Mandel}},
  \bibinfo {author} {\bibfnamefont {T.}~\bibnamefont {Esslinger}}, \bibinfo
  {author} {\bibfnamefont {T.~W.}\ \bibnamefont {H{\"a}nsch}}, \ and\ \bibinfo
  {author} {\bibfnamefont {I.}~\bibnamefont {Bloch}},\ }\href {\doibase
  10.1038/415039a} {\bibfield  {journal} {\bibinfo  {journal} {Nature}\
  }\textbf {\bibinfo {volume} {415}},\ \bibinfo {pages} {39} (\bibinfo {year}
  {2002})}\BibitemShut {NoStop}%
\bibitem [{\citenamefont {Duan}\ \emph {et~al.}(2003)\citenamefont {Duan},
  \citenamefont {Demler},\ and\ \citenamefont {Lukin}}]{Duan2003}%
  \BibitemOpen
  \bibfield  {author} {\bibinfo {author} {\bibfnamefont {L.-M.}\ \bibnamefont
  {Duan}}, \bibinfo {author} {\bibfnamefont {E.}~\bibnamefont {Demler}}, \ and\
  \bibinfo {author} {\bibfnamefont {M.~D.}\ \bibnamefont {Lukin}},\ }\href
  {\doibase 10.1103/PhysRevLett.91.090402} {\bibfield  {journal} {\bibinfo
  {journal} {Phys. Rev. Lett.}\ }\textbf {\bibinfo {volume} {91}},\ \bibinfo
  {pages} {090402} (\bibinfo {year} {2003})}\BibitemShut {NoStop}%
\bibitem [{\citenamefont {Bloch}(2005)}]{Bloch_2005}%
  \BibitemOpen
  \bibfield  {author} {\bibinfo {author} {\bibfnamefont {I.}~\bibnamefont
  {Bloch}},\ }\href {\doibase 10.1088/0953-4075/38/9/013} {\bibfield  {journal}
  {\bibinfo  {journal} {Journal of Physics B: Atomic, Molecular and Optical
  Physics}\ }\textbf {\bibinfo {volume} {38}},\ \bibinfo {pages} {S629}
  (\bibinfo {year} {2005})}\BibitemShut {NoStop}%
\bibitem [{\citenamefont {Bloch}\ \emph {et~al.}(2008)\citenamefont {Bloch},
  \citenamefont {Dalibard},\ and\ \citenamefont {Zwerger}}]{Bloch2008}%
  \BibitemOpen
  \bibfield  {author} {\bibinfo {author} {\bibfnamefont {I.}~\bibnamefont
  {Bloch}}, \bibinfo {author} {\bibfnamefont {J.}~\bibnamefont {Dalibard}}, \
  and\ \bibinfo {author} {\bibfnamefont {W.}~\bibnamefont {Zwerger}},\ }\href
  {\doibase 10.1103/RevModPhys.80.885} {\bibfield  {journal} {\bibinfo
  {journal} {Rev. Mod. Phys.}\ }\textbf {\bibinfo {volume} {80}},\ \bibinfo
  {pages} {885} (\bibinfo {year} {2008})}\BibitemShut {NoStop}%
\bibitem [{\citenamefont {Struck}\ \emph {et~al.}(2013)\citenamefont {Struck},
  \citenamefont {Weinberg}, \citenamefont {{\"O}lschl{\"a}ger}, \citenamefont
  {Windpassinger}, \citenamefont {Simonet}, \citenamefont {Sengstock},
  \citenamefont {H{\"o}ppner}, \citenamefont {Hauke}, \citenamefont {Eckardt},
  \citenamefont {Lewenstein},\ and\ \citenamefont {Mathey}}]{Struck2013}%
  \BibitemOpen
  \bibfield  {author} {\bibinfo {author} {\bibfnamefont {J.}~\bibnamefont
  {Struck}}, \bibinfo {author} {\bibfnamefont {M.}~\bibnamefont {Weinberg}},
  \bibinfo {author} {\bibfnamefont {C.}~\bibnamefont {{\"O}lschl{\"a}ger}},
  \bibinfo {author} {\bibfnamefont {P.}~\bibnamefont {Windpassinger}}, \bibinfo
  {author} {\bibfnamefont {J.}~\bibnamefont {Simonet}}, \bibinfo {author}
  {\bibfnamefont {K.}~\bibnamefont {Sengstock}}, \bibinfo {author}
  {\bibfnamefont {R.}~\bibnamefont {H{\"o}ppner}}, \bibinfo {author}
  {\bibfnamefont {P.}~\bibnamefont {Hauke}}, \bibinfo {author} {\bibfnamefont
  {A.}~\bibnamefont {Eckardt}}, \bibinfo {author} {\bibfnamefont
  {M.}~\bibnamefont {Lewenstein}}, \ and\ \bibinfo {author} {\bibfnamefont
  {L.}~\bibnamefont {Mathey}},\ }\href {\doibase 10.1038/nphys2750} {\bibfield
  {journal} {\bibinfo  {journal} {Nature Physics}\ }\textbf {\bibinfo {volume}
  {9}},\ \bibinfo {pages} {738} (\bibinfo {year} {2013})}\BibitemShut {NoStop}%
\bibitem [{\citenamefont {Burgarth}\ and\ \citenamefont
  {Bose}(2005{\natexlab{a}})}]{Burgarth2005}%
  \BibitemOpen
  \bibfield  {author} {\bibinfo {author} {\bibfnamefont {D.}~\bibnamefont
  {Burgarth}}\ and\ \bibinfo {author} {\bibfnamefont {S.}~\bibnamefont
  {Bose}},\ }\href {\doibase 10.1103/PhysRevA.71.052315} {\bibfield  {journal}
  {\bibinfo  {journal} {Phys. Rev. A}\ }\textbf {\bibinfo {volume} {71}},\
  \bibinfo {pages} {052315} (\bibinfo {year} {2005}{\natexlab{a}})}\BibitemShut
  {NoStop}%
\bibitem [{\citenamefont {Burgarth}\ \emph {et~al.}(2005)\citenamefont
  {Burgarth}, \citenamefont {Giovannetti},\ and\ \citenamefont
  {Bose}}]{Burgarth2005a}%
  \BibitemOpen
  \bibfield  {author} {\bibinfo {author} {\bibfnamefont {D.}~\bibnamefont
  {Burgarth}}, \bibinfo {author} {\bibfnamefont {V.}~\bibnamefont
  {Giovannetti}}, \ and\ \bibinfo {author} {\bibfnamefont {S.}~\bibnamefont
  {Bose}},\ }\href {\doibase 10.1088/0305-4470/38/30/013} {\bibfield  {journal}
  {\bibinfo  {journal} {Journal of Physics A: Mathematical and General}\
  }\textbf {\bibinfo {volume} {38}},\ \bibinfo {pages} {6793} (\bibinfo {year}
  {2005})}\BibitemShut {NoStop}%
\bibitem [{\citenamefont {Burgarth}\ and\ \citenamefont
  {Bose}(2005{\natexlab{b}})}]{Burgarth2005b}%
  \BibitemOpen
  \bibfield  {author} {\bibinfo {author} {\bibfnamefont {D.}~\bibnamefont
  {Burgarth}}\ and\ \bibinfo {author} {\bibfnamefont {S.}~\bibnamefont
  {Bose}},\ }\href {\doibase 10.1088/1367-2630/7/1/135} {\bibfield  {journal}
  {\bibinfo  {journal} {New Journal of Physics}\ }\textbf {\bibinfo {volume}
  {7}},\ \bibinfo {pages} {135} (\bibinfo {year}
  {2005}{\natexlab{b}})}\BibitemShut {NoStop}%
\bibitem [{\citenamefont {Vaucher}\ \emph {et~al.}(2005)\citenamefont
  {Vaucher}, \citenamefont {Burgarth},\ and\ \citenamefont
  {Bose}}]{Vaucher_2005}%
  \BibitemOpen
  \bibfield  {author} {\bibinfo {author} {\bibfnamefont {B.}~\bibnamefont
  {Vaucher}}, \bibinfo {author} {\bibfnamefont {D.}~\bibnamefont {Burgarth}}, \
  and\ \bibinfo {author} {\bibfnamefont {S.}~\bibnamefont {Bose}},\ }\href
  {\doibase 10.1088/1464-4266/7/10/023} {\bibfield  {journal} {\bibinfo
  {journal} {Journal of Optics B: Quantum and Semiclassical Optics}\ }\textbf
  {\bibinfo {volume} {7}},\ \bibinfo {pages} {S356} (\bibinfo {year}
  {2005})}\BibitemShut {NoStop}%
\bibitem [{\citenamefont {Li}\ \emph {et~al.}(2005)\citenamefont {Li},
  \citenamefont {Shi}, \citenamefont {Chen}, \citenamefont {Song},\ and\
  \citenamefont {Sun}}]{Li2005}%
  \BibitemOpen
  \bibfield  {author} {\bibinfo {author} {\bibfnamefont {Y.}~\bibnamefont
  {Li}}, \bibinfo {author} {\bibfnamefont {T.}~\bibnamefont {Shi}}, \bibinfo
  {author} {\bibfnamefont {B.}~\bibnamefont {Chen}}, \bibinfo {author}
  {\bibfnamefont {Z.}~\bibnamefont {Song}}, \ and\ \bibinfo {author}
  {\bibfnamefont {C.-P.}\ \bibnamefont {Sun}},\ }\href {\doibase
  10.1103/PhysRevA.71.022301} {\bibfield  {journal} {\bibinfo  {journal} {Phys.
  Rev. A}\ }\textbf {\bibinfo {volume} {71}},\ \bibinfo {pages} {022301}
  (\bibinfo {year} {2005})}\BibitemShut {NoStop}%
\bibitem [{\citenamefont {Almeida}\ \emph {et~al.}(2019)\citenamefont
  {Almeida}, \citenamefont {Souza}, \citenamefont {{de Moura}},\ and\
  \citenamefont {Lyra}}]{Almeida2019}%
  \BibitemOpen
  \bibfield  {author} {\bibinfo {author} {\bibfnamefont {G.~M.}\ \bibnamefont
  {Almeida}}, \bibinfo {author} {\bibfnamefont {A.~M.}\ \bibnamefont {Souza}},
  \bibinfo {author} {\bibfnamefont {F.~A.}\ \bibnamefont {{de Moura}}}, \ and\
  \bibinfo {author} {\bibfnamefont {M.~L.}\ \bibnamefont {Lyra}},\ }\href
  {\doibase https://doi.org/10.1016/j.physleta.2019.125847} {\bibfield
  {journal} {\bibinfo  {journal} {Physics Letters A}\ }\textbf {\bibinfo
  {volume} {383}},\ \bibinfo {pages} {125847} (\bibinfo {year}
  {2019})}\BibitemShut {NoStop}%
\bibitem [{\citenamefont {KAY}(2010)}]{Kay2010}%
  \BibitemOpen
  \bibfield  {author} {\bibinfo {author} {\bibfnamefont {A.}~\bibnamefont
  {KAY}},\ }\href {\doibase 10.1142/S0219749910006514} {\bibfield  {journal}
  {\bibinfo  {journal} {International Journal of Quantum Information}\ }\textbf
  {\bibinfo {volume} {08}},\ \bibinfo {pages} {641} (\bibinfo {year}
  {2010})}\BibitemShut {NoStop}%
\bibitem [{\citenamefont {Yao}\ \emph {et~al.}(2011)\citenamefont {Yao},
  \citenamefont {Jiang}, \citenamefont {Gorshkov}, \citenamefont {Gong},
  \citenamefont {Zhai}, \citenamefont {Duan},\ and\ \citenamefont
  {Lukin}}]{Yao2011}%
  \BibitemOpen
  \bibfield  {author} {\bibinfo {author} {\bibfnamefont {N.~Y.}\ \bibnamefont
  {Yao}}, \bibinfo {author} {\bibfnamefont {L.}~\bibnamefont {Jiang}}, \bibinfo
  {author} {\bibfnamefont {A.~V.}\ \bibnamefont {Gorshkov}}, \bibinfo {author}
  {\bibfnamefont {Z.-X.}\ \bibnamefont {Gong}}, \bibinfo {author}
  {\bibfnamefont {A.}~\bibnamefont {Zhai}}, \bibinfo {author} {\bibfnamefont
  {L.-M.}\ \bibnamefont {Duan}}, \ and\ \bibinfo {author} {\bibfnamefont
  {M.~D.}\ \bibnamefont {Lukin}},\ }\href {\doibase
  10.1103/PhysRevLett.106.040505} {\bibfield  {journal} {\bibinfo  {journal}
  {Phys. Rev. Lett.}\ }\textbf {\bibinfo {volume} {106}},\ \bibinfo {pages}
  {040505} (\bibinfo {year} {2011})}\BibitemShut {NoStop}%
\bibitem [{\citenamefont {{Acosta Coden}}\ \emph {et~al.}(2021)\citenamefont
  {{Acosta Coden}}, \citenamefont {Gómez}, \citenamefont {Ferrón},\ and\
  \citenamefont {Osenda}}]{ACOSTACODEN2021}%
  \BibitemOpen
  \bibfield  {author} {\bibinfo {author} {\bibfnamefont {D.}~\bibnamefont
  {{Acosta Coden}}}, \bibinfo {author} {\bibfnamefont {S.}~\bibnamefont
  {Gómez}}, \bibinfo {author} {\bibfnamefont {A.}~\bibnamefont {Ferrón}}, \
  and\ \bibinfo {author} {\bibfnamefont {O.}~\bibnamefont {Osenda}},\
  }\href@noop {} {\bibfield  {journal} {\bibinfo  {journal} {Physics Letters
  A}\ }\textbf {\bibinfo {volume} {387}},\ \bibinfo {pages} {127009} (\bibinfo
  {year} {2021})}\BibitemShut {NoStop}%
\bibitem [{\citenamefont {Christandl}\ \emph {et~al.}(2004)\citenamefont
  {Christandl}, \citenamefont {Datta}, \citenamefont {Ekert},\ and\
  \citenamefont {Landahl}}]{Christandl2004}%
  \BibitemOpen
  \bibfield  {author} {\bibinfo {author} {\bibfnamefont {M.}~\bibnamefont
  {Christandl}}, \bibinfo {author} {\bibfnamefont {N.}~\bibnamefont {Datta}},
  \bibinfo {author} {\bibfnamefont {A.}~\bibnamefont {Ekert}}, \ and\ \bibinfo
  {author} {\bibfnamefont {A.~J.}\ \bibnamefont {Landahl}},\ }\href {\doibase
  10.1103/PhysRevLett.92.187902} {\bibfield  {journal} {\bibinfo  {journal}
  {Phys. Rev. Lett.}\ }\textbf {\bibinfo {volume} {92}},\ \bibinfo {pages}
  {187902} (\bibinfo {year} {2004})}\BibitemShut {NoStop}%
\bibitem [{\citenamefont {Banchi}(2013)}]{Banchi2013}%
  \BibitemOpen
  \bibfield  {author} {\bibinfo {author} {\bibfnamefont {L.}~\bibnamefont
  {Banchi}},\ }\href {\doibase 10.1140/epjp/i2013-13137-6} {\bibfield
  {journal} {\bibinfo  {journal} {The European Physical Journal Plus}\ }\textbf
  {\bibinfo {volume} {128}},\ \bibinfo {pages} {137} (\bibinfo {year}
  {2013})}\BibitemShut {NoStop}%
\bibitem [{\citenamefont {Karbach}\ and\ \citenamefont
  {Stolze}(2005)}]{Karbach2005}%
  \BibitemOpen
  \bibfield  {author} {\bibinfo {author} {\bibfnamefont {P.}~\bibnamefont
  {Karbach}}\ and\ \bibinfo {author} {\bibfnamefont {J.}~\bibnamefont
  {Stolze}},\ }\href {\doibase 10.1103/PhysRevA.72.030301} {\bibfield
  {journal} {\bibinfo  {journal} {Phys. Rev. A}\ }\textbf {\bibinfo {volume}
  {72}},\ \bibinfo {pages} {030301} (\bibinfo {year} {2005})}\BibitemShut
  {NoStop}%
\bibitem [{\citenamefont {Subrahmanyam}(2004)}]{Subrahmanyam2004}%
  \BibitemOpen
  \bibfield  {author} {\bibinfo {author} {\bibfnamefont {V.}~\bibnamefont
  {Subrahmanyam}},\ }\href {\doibase 10.1103/PhysRevA.69.034304} {\bibfield
  {journal} {\bibinfo  {journal} {Phys. Rev. A}\ }\textbf {\bibinfo {volume}
  {69}},\ \bibinfo {pages} {034304} (\bibinfo {year} {2004})}\BibitemShut
  {NoStop}%
\bibitem [{\citenamefont {Liu}\ \emph {et~al.}(2008)\citenamefont {Liu},
  \citenamefont {Zhang},\ and\ \citenamefont {Chen}}]{LIU2008}%
  \BibitemOpen
  \bibfield  {author} {\bibinfo {author} {\bibfnamefont {J.}~\bibnamefont
  {Liu}}, \bibinfo {author} {\bibfnamefont {G.-F.}\ \bibnamefont {Zhang}}, \
  and\ \bibinfo {author} {\bibfnamefont {Z.-Y.}\ \bibnamefont {Chen}},\ }\href
  {\doibase https://doi.org/10.1016/j.physleta.2008.01.017} {\bibfield
  {journal} {\bibinfo  {journal} {Physics Letters A}\ }\textbf {\bibinfo
  {volume} {372}},\ \bibinfo {pages} {2830} (\bibinfo {year}
  {2008})}\BibitemShut {NoStop}%
\bibitem [{\citenamefont {Fel'dman}\ and\ \citenamefont
  {Zenchuk}(2009)}]{FELDMAN20091719}%
  \BibitemOpen
  \bibfield  {author} {\bibinfo {author} {\bibfnamefont {E.}~\bibnamefont
  {Fel'dman}}\ and\ \bibinfo {author} {\bibfnamefont {A.}~\bibnamefont
  {Zenchuk}},\ }\href {\doibase https://doi.org/10.1016/j.physleta.2009.03.038}
  {\bibfield  {journal} {\bibinfo  {journal} {Physics Letters A}\ }\textbf
  {\bibinfo {volume} {373}},\ \bibinfo {pages} {1719} (\bibinfo {year}
  {2009})}\BibitemShut {NoStop}%
\bibitem [{\citenamefont {Pouyandeh}\ and\ \citenamefont
  {Shahbazi}(2015)}]{Pouyandeh2015}%
  \BibitemOpen
  \bibfield  {author} {\bibinfo {author} {\bibfnamefont {S.}~\bibnamefont
  {Pouyandeh}}\ and\ \bibinfo {author} {\bibfnamefont {F.}~\bibnamefont
  {Shahbazi}},\ }\href {\doibase 10.1142/S0219749915500306} {\bibfield
  {journal} {\bibinfo  {journal} {International Journal of Quantum
  Information}\ }\textbf {\bibinfo {volume} {13}},\ \bibinfo {pages} {1550030}
  (\bibinfo {year} {2015})}\BibitemShut {NoStop}%
\bibitem [{\citenamefont {Yang}\ \emph {et~al.}(2015)\citenamefont {Yang},
  \citenamefont {Gao},\ and\ \citenamefont {Qin}}]{Yang2015}%
  \BibitemOpen
  \bibfield  {author} {\bibinfo {author} {\bibfnamefont {Z.}~\bibnamefont
  {Yang}}, \bibinfo {author} {\bibfnamefont {M.}~\bibnamefont {Gao}}, \ and\
  \bibinfo {author} {\bibfnamefont {W.}~\bibnamefont {Qin}},\ }\href {\doibase
  10.1142/S0217979215502070} {\bibfield  {journal} {\bibinfo  {journal}
  {International Journal of Modern Physics B}\ }\textbf {\bibinfo {volume}
  {29}},\ \bibinfo {pages} {1550207} (\bibinfo {year} {2015})}\BibitemShut
  {NoStop}%
\bibitem [{\citenamefont {Shan}\ \emph {et~al.}(2018)\citenamefont {Shan},
  \citenamefont {Dai}, \citenamefont {Shen},\ and\ \citenamefont
  {Yi}}]{Shan2018}%
  \BibitemOpen
  \bibfield  {author} {\bibinfo {author} {\bibfnamefont {H.~J.}\ \bibnamefont
  {Shan}}, \bibinfo {author} {\bibfnamefont {C.~M.}\ \bibnamefont {Dai}},
  \bibinfo {author} {\bibfnamefont {H.~Z.}\ \bibnamefont {Shen}}, \ and\
  \bibinfo {author} {\bibfnamefont {X.~X.}\ \bibnamefont {Yi}},\ }\href
  {\doibase 10.1038/s41598-018-31552-w} {\bibfield  {journal} {\bibinfo
  {journal} {Scientific Reports}\ }\textbf {\bibinfo {volume} {8}},\ \bibinfo
  {pages} {13565} (\bibinfo {year} {2018})}\BibitemShut {NoStop}%
\bibitem [{\citenamefont {Almeida}\ \emph {et~al.}(2018)\citenamefont
  {Almeida}, \citenamefont {{de Moura}},\ and\ \citenamefont
  {Lyra}}]{ALMEIDA2018}%
  \BibitemOpen
  \bibfield  {author} {\bibinfo {author} {\bibfnamefont {G.~M.}\ \bibnamefont
  {Almeida}}, \bibinfo {author} {\bibfnamefont {F.~A.}\ \bibnamefont {{de
  Moura}}}, \ and\ \bibinfo {author} {\bibfnamefont {M.~L.}\ \bibnamefont
  {Lyra}},\ }\href {\doibase https://doi.org/10.1016/j.physleta.2018.03.028}
  {\bibfield  {journal} {\bibinfo  {journal} {Physics Letters A}\ }\textbf
  {\bibinfo {volume} {382}},\ \bibinfo {pages} {1335} (\bibinfo {year}
  {2018})}\BibitemShut {NoStop}%
\bibitem [{\citenamefont {Hermes}\ \emph {et~al.}(2020)\citenamefont {Hermes},
  \citenamefont {Apollaro}, \citenamefont {Paganelli},\ and\ \citenamefont
  {Macr\`{\i}}}]{Hermes2020}%
  \BibitemOpen
  \bibfield  {author} {\bibinfo {author} {\bibfnamefont {S.}~\bibnamefont
  {Hermes}}, \bibinfo {author} {\bibfnamefont {T.~J.~G.}\ \bibnamefont
  {Apollaro}}, \bibinfo {author} {\bibfnamefont {S.}~\bibnamefont {Paganelli}},
  \ and\ \bibinfo {author} {\bibfnamefont {T.}~\bibnamefont {Macr\`{\i}}},\
  }\href {\doibase 10.1103/PhysRevA.101.053607} {\bibfield  {journal} {\bibinfo
   {journal} {Phys. Rev. A}\ }\textbf {\bibinfo {volume} {101}},\ \bibinfo
  {pages} {053607} (\bibinfo {year} {2020})}\BibitemShut {NoStop}%
\bibitem [{\citenamefont {Ronke}\ \emph {et~al.}(2011)\citenamefont {Ronke},
  \citenamefont {Spiller},\ and\ \citenamefont {D'Amico}}]{Ronke2011}%
  \BibitemOpen
  \bibfield  {author} {\bibinfo {author} {\bibfnamefont {R.}~\bibnamefont
  {Ronke}}, \bibinfo {author} {\bibfnamefont {T.~P.}\ \bibnamefont {Spiller}},
  \ and\ \bibinfo {author} {\bibfnamefont {I.}~\bibnamefont {D'Amico}},\ }\href
  {\doibase 10.1103/PhysRevA.83.012325} {\bibfield  {journal} {\bibinfo
  {journal} {Phys. Rev. A}\ }\textbf {\bibinfo {volume} {83}},\ \bibinfo
  {pages} {012325} (\bibinfo {year} {2011})}\BibitemShut {NoStop}%
\bibitem [{\citenamefont {Apollaro}\ \emph {et~al.}(2023)\citenamefont
  {Apollaro}, \citenamefont {Lorenzo}, \citenamefont {Plastina}, \citenamefont
  {Consiglio},\ and\ \citenamefont {Życzkowski}}]{Apollaro2023}%
  \BibitemOpen
  \bibfield  {author} {\bibinfo {author} {\bibfnamefont {T.~J.~G.}\
  \bibnamefont {Apollaro}}, \bibinfo {author} {\bibfnamefont {S.}~\bibnamefont
  {Lorenzo}}, \bibinfo {author} {\bibfnamefont {F.}~\bibnamefont {Plastina}},
  \bibinfo {author} {\bibfnamefont {M.}~\bibnamefont {Consiglio}}, \ and\
  \bibinfo {author} {\bibfnamefont {K.}~\bibnamefont {Życzkowski}},\ }\href
  {\doibase 10.3390/e25010046} {\bibfield  {journal} {\bibinfo  {journal}
  {Entropy}\ }\textbf {\bibinfo {volume} {25}} (\bibinfo {year} {2023}),\
  10.3390/e25010046}\BibitemShut {NoStop}%
\bibitem [{\citenamefont {Chapman}\ \emph {et~al.}(2016)\citenamefont
  {Chapman}, \citenamefont {Santandrea}, \citenamefont {Huang}, \citenamefont
  {Corrielli}, \citenamefont {Crespi}, \citenamefont {Yung}, \citenamefont
  {Osellame},\ and\ \citenamefont {Peruzzo}}]{chapman2015}%
  \BibitemOpen
  \bibfield  {author} {\bibinfo {author} {\bibfnamefont {R.~J.}\ \bibnamefont
  {Chapman}}, \bibinfo {author} {\bibfnamefont {M.}~\bibnamefont {Santandrea}},
  \bibinfo {author} {\bibfnamefont {Z.}~\bibnamefont {Huang}}, \bibinfo
  {author} {\bibfnamefont {G.}~\bibnamefont {Corrielli}}, \bibinfo {author}
  {\bibfnamefont {A.}~\bibnamefont {Crespi}}, \bibinfo {author} {\bibfnamefont
  {M.-H.}\ \bibnamefont {Yung}}, \bibinfo {author} {\bibfnamefont
  {R.}~\bibnamefont {Osellame}}, \ and\ \bibinfo {author} {\bibfnamefont
  {A.}~\bibnamefont {Peruzzo}},\ }\href {\doibase 10.1038/ncomms11339}
  {\bibfield  {journal} {\bibinfo  {journal} {Nature Communications}\ }\textbf
  {\bibinfo {volume} {7}},\ \bibinfo {pages} {11339} (\bibinfo {year}
  {2016})}\BibitemShut {NoStop}%
\bibitem [{\citenamefont {Sousa}\ and\ \citenamefont {Omar}(2014)}]{Sousa2014}%
  \BibitemOpen
  \bibfield  {author} {\bibinfo {author} {\bibfnamefont {R.}~\bibnamefont
  {Sousa}}\ and\ \bibinfo {author} {\bibfnamefont {Y.}~\bibnamefont {Omar}},\
  }\href {\doibase 10.1088/1367-2630/16/12/123003} {\bibfield  {journal}
  {\bibinfo  {journal} {New Journal of Physics}\ }\textbf {\bibinfo {volume}
  {16}},\ \bibinfo {pages} {123003} (\bibinfo {year} {2014})}\BibitemShut
  {NoStop}%
\bibitem [{\citenamefont {Banchi}\ \emph {et~al.}(2017)\citenamefont {Banchi},
  \citenamefont {Coutinho}, \citenamefont {Godsil},\ and\ \citenamefont
  {Severini}}]{Banchi2017}%
  \BibitemOpen
  \bibfield  {author} {\bibinfo {author} {\bibfnamefont {L.}~\bibnamefont
  {Banchi}}, \bibinfo {author} {\bibfnamefont {G.}~\bibnamefont {Coutinho}},
  \bibinfo {author} {\bibfnamefont {C.}~\bibnamefont {Godsil}}, \ and\ \bibinfo
  {author} {\bibfnamefont {S.}~\bibnamefont {Severini}},\ }\href {\doibase
  10.1063/1.4978327} {\bibfield  {journal} {\bibinfo  {journal} {Journal of
  Mathematical Physics}\ }\textbf {\bibinfo {volume} {58}},\ \bibinfo {pages}
  {032202} (\bibinfo {year} {2017})}\BibitemShut {NoStop}%
\bibitem [{\citenamefont {Serra}\ \emph {et~al.}(2022)\citenamefont {Serra},
  \citenamefont {Ferrón},\ and\ \citenamefont {Osenda}}]{Serra2022}%
  \BibitemOpen
  \bibfield  {author} {\bibinfo {author} {\bibfnamefont {P.}~\bibnamefont
  {Serra}}, \bibinfo {author} {\bibfnamefont {A.}~\bibnamefont {Ferrón}}, \
  and\ \bibinfo {author} {\bibfnamefont {O.}~\bibnamefont {Osenda}},\ }\href
  {\doibase 10.1088/1751-8121/ac901d} {\bibfield  {journal} {\bibinfo
  {journal} {Journal of Physics A: Mathematical and Theoretical}\ }\textbf
  {\bibinfo {volume} {55}},\ \bibinfo {pages} {405302} (\bibinfo {year}
  {2022})}\BibitemShut {NoStop}%
\bibitem [{\citenamefont {Ercolessi}(2003)}]{Ercolessi2003}%
  \BibitemOpen
  \bibfield  {author} {\bibinfo {author} {\bibfnamefont {E.}~\bibnamefont
  {Ercolessi}},\ }\href {\doibase 10.1142/S0217732303012544} {\bibfield
  {journal} {\bibinfo  {journal} {Modern Physics Letters A}\ }\textbf {\bibinfo
  {volume} {18}},\ \bibinfo {pages} {2329} (\bibinfo {year}
  {2003})}\BibitemShut {NoStop}%
\bibitem [{\citenamefont {Ivanov}(2009)}]{Ivanov2009}%
  \BibitemOpen
  \bibfield  {author} {\bibinfo {author} {\bibfnamefont {N.}~\bibnamefont
  {Ivanov}},\ }\href {\doibase 10.48550/arXiv.0909.2182} {\bibfield  {journal}
  {\bibinfo  {journal} {arXiv:0909.2182}\ } (\bibinfo {year} {2009}),\
  10.48550/arXiv.0909.2182}\BibitemShut {NoStop}%
\bibitem [{\citenamefont {Dagotto}\ and\ \citenamefont
  {Rice}(1996)}]{Dagotto1996}%
  \BibitemOpen
  \bibfield  {author} {\bibinfo {author} {\bibfnamefont {E.}~\bibnamefont
  {Dagotto}}\ and\ \bibinfo {author} {\bibfnamefont {T.~M.}\ \bibnamefont
  {Rice}},\ }\href {\doibase 10.1126/science.271.5249.618} {\bibfield
  {journal} {\bibinfo  {journal} {Science}\ }\textbf {\bibinfo {volume}
  {271}},\ \bibinfo {pages} {618} (\bibinfo {year} {1996})}\BibitemShut
  {NoStop}%
\bibitem [{\citenamefont {Batchelor}\ \emph
  {et~al.}(2003{\natexlab{a}})\citenamefont {Batchelor}, \citenamefont {Guan},
  \citenamefont {Foerster},\ and\ \citenamefont {Zhou}}]{Batchelor2003}%
  \BibitemOpen
  \bibfield  {author} {\bibinfo {author} {\bibfnamefont {M.~T.}\ \bibnamefont
  {Batchelor}}, \bibinfo {author} {\bibfnamefont {X.-W.}\ \bibnamefont {Guan}},
  \bibinfo {author} {\bibfnamefont {A.}~\bibnamefont {Foerster}}, \ and\
  \bibinfo {author} {\bibfnamefont {H.-Q.}\ \bibnamefont {Zhou}},\ }\href
  {\doibase 10.1088/1367-2630/5/1/107} {\bibfield  {journal} {\bibinfo
  {journal} {New Journal of Physics}\ }\textbf {\bibinfo {volume} {5}},\
  \bibinfo {pages} {107} (\bibinfo {year} {2003}{\natexlab{a}})}\BibitemShut
  {NoStop}%
\bibitem [{\citenamefont {Batchelor}\ \emph
  {et~al.}(2003{\natexlab{b}})\citenamefont {Batchelor}, \citenamefont {Guan},
  \citenamefont {Foerster}, \citenamefont {Tonel},\ and\ \citenamefont
  {Zhou}}]{Batchelor2003a}%
  \BibitemOpen
  \bibfield  {author} {\bibinfo {author} {\bibfnamefont {M.}~\bibnamefont
  {Batchelor}}, \bibinfo {author} {\bibfnamefont {X.-W.}\ \bibnamefont {Guan}},
  \bibinfo {author} {\bibfnamefont {A.}~\bibnamefont {Foerster}}, \bibinfo
  {author} {\bibfnamefont {A.}~\bibnamefont {Tonel}}, \ and\ \bibinfo {author}
  {\bibfnamefont {H.-Q.}\ \bibnamefont {Zhou}},\ }\href {\doibase
  https://doi.org/10.1016/j.nuclphysb.2003.07.012} {\bibfield  {journal}
  {\bibinfo  {journal} {Nuclear Physics B}\ }\textbf {\bibinfo {volume}
  {669}},\ \bibinfo {pages} {385} (\bibinfo {year}
  {2003}{\natexlab{b}})}\BibitemShut {NoStop}%
\bibitem [{\citenamefont {Batchelor}\ \emph {et~al.}(2007)\citenamefont
  {Batchelor}, \citenamefont {Guan}, \citenamefont {Oelkers},\ and\
  \citenamefont {Tsuboi}}]{Batchelor2007}%
  \BibitemOpen
  \bibfield  {author} {\bibinfo {author} {\bibfnamefont {M.~T.}\ \bibnamefont
  {Batchelor}}, \bibinfo {author} {\bibfnamefont {X.~W.}\ \bibnamefont {Guan}},
  \bibinfo {author} {\bibfnamefont {N.}~\bibnamefont {Oelkers}}, \ and\
  \bibinfo {author} {\bibfnamefont {Z.}~\bibnamefont {Tsuboi}},\ }\href
  {\doibase 10.1080/00018730701265383} {\bibfield  {journal} {\bibinfo
  {journal} {Advances in Physics}\ }\textbf {\bibinfo {volume} {56}},\ \bibinfo
  {pages} {465} (\bibinfo {year} {2007})}\BibitemShut {NoStop}%
\bibitem [{\citenamefont {Miki}\ \emph {et~al.}(2012)\citenamefont {Miki},
  \citenamefont {Tsujimoto}, \citenamefont {Vinet},\ and\ \citenamefont
  {Zhedanov}}]{Miki2012}%
  \BibitemOpen
  \bibfield  {author} {\bibinfo {author} {\bibfnamefont {H.}~\bibnamefont
  {Miki}}, \bibinfo {author} {\bibfnamefont {S.}~\bibnamefont {Tsujimoto}},
  \bibinfo {author} {\bibfnamefont {L.}~\bibnamefont {Vinet}}, \ and\ \bibinfo
  {author} {\bibfnamefont {A.}~\bibnamefont {Zhedanov}},\ }\href {\doibase
  10.1103/PhysRevA.85.062306} {\bibfield  {journal} {\bibinfo  {journal} {Phys.
  Rev. A}\ }\textbf {\bibinfo {volume} {85}},\ \bibinfo {pages} {062306}
  (\bibinfo {year} {2012})}\BibitemShut {NoStop}%
\bibitem [{\citenamefont {Post}(2015)}]{Post2015}%
  \BibitemOpen
  \bibfield  {author} {\bibinfo {author} {\bibfnamefont {S.}~\bibnamefont
  {Post}},\ }\href {\doibase 10.1007/s10440-014-9953-5} {\bibfield  {journal}
  {\bibinfo  {journal} {Acta Applicandae Mathematicae}\ }\textbf {\bibinfo
  {volume} {135}},\ \bibinfo {pages} {209} (\bibinfo {year}
  {2015})}\BibitemShut {NoStop}%
\bibitem [{\citenamefont {Heisenberg}(1928)}]{Heisenberg1928}%
  \BibitemOpen
  \bibfield  {author} {\bibinfo {author} {\bibfnamefont {W.}~\bibnamefont
  {Heisenberg}},\ }\href {\doibase 10.1007/BF01328601} {\bibfield  {journal}
  {\bibinfo  {journal} {Zeitschrift f{\"u}r Physik}\ }\textbf {\bibinfo
  {volume} {49}},\ \bibinfo {pages} {619} (\bibinfo {year} {1928})}\BibitemShut
  {NoStop}%
\bibitem [{\citenamefont {Okwamoto}(1984)}]{Okwamoto1984}%
  \BibitemOpen
  \bibfield  {author} {\bibinfo {author} {\bibfnamefont {Y.}~\bibnamefont
  {Okwamoto}},\ }\href {\doibase 10.1143/JPSJ.53.2434} {\bibfield  {journal}
  {\bibinfo  {journal} {Journal of the Physical Society of Japan}\ }\textbf
  {\bibinfo {volume} {53}},\ \bibinfo {pages} {2434} (\bibinfo {year}
  {1984})}\BibitemShut {NoStop}%
\bibitem [{\citenamefont {Aplesnin}(1999)}]{Aplesnin1999}%
  \BibitemOpen
  \bibfield  {author} {\bibinfo {author} {\bibfnamefont {S.~S.}\ \bibnamefont
  {Aplesnin}},\ }\href {\doibase 10.1134/1.1130737} {\bibfield  {journal}
  {\bibinfo  {journal} {Physics of the Solid State}\ }\textbf {\bibinfo
  {volume} {41}},\ \bibinfo {pages} {103} (\bibinfo {year} {1999})}\BibitemShut
  {NoStop}%
\bibitem [{\citenamefont {Weihong}\ \emph {et~al.}(1999)\citenamefont
  {Weihong}, \citenamefont {McKenzie},\ and\ \citenamefont
  {Singh}}]{Zheng1999}%
  \BibitemOpen
  \bibfield  {author} {\bibinfo {author} {\bibfnamefont {Z.}~\bibnamefont
  {Weihong}}, \bibinfo {author} {\bibfnamefont {R.~H.}\ \bibnamefont
  {McKenzie}}, \ and\ \bibinfo {author} {\bibfnamefont {R.~R.~P.}\ \bibnamefont
  {Singh}},\ }\href {\doibase 10.1103/PhysRevB.59.14367} {\bibfield  {journal}
  {\bibinfo  {journal} {Phys. Rev. B}\ }\textbf {\bibinfo {volume} {59}},\
  \bibinfo {pages} {14367} (\bibinfo {year} {1999})}\BibitemShut {NoStop}%
\bibitem [{\citenamefont {Costa}\ and\ \citenamefont
  {Pires}(2003)}]{Costa2003}%
  \BibitemOpen
  \bibfield  {author} {\bibinfo {author} {\bibfnamefont {B.}~\bibnamefont
  {Costa}}\ and\ \bibinfo {author} {\bibfnamefont {A.}~\bibnamefont {Pires}},\
  }\href {\doibase https://doi.org/10.1016/S0304-8853(02)01527-5} {\bibfield
  {journal} {\bibinfo  {journal} {Journal of Magnetism and Magnetic Materials}\
  }\textbf {\bibinfo {volume} {262}},\ \bibinfo {pages} {316} (\bibinfo {year}
  {2003})}\BibitemShut {NoStop}%
\bibitem [{\citenamefont {Cuccoli}\ \emph {et~al.}(2006)\citenamefont
  {Cuccoli}, \citenamefont {Gori}, \citenamefont {Vaia},\ and\ \citenamefont
  {Verrucchi}}]{Cuccoli2006}%
  \BibitemOpen
  \bibfield  {author} {\bibinfo {author} {\bibfnamefont {A.}~\bibnamefont
  {Cuccoli}}, \bibinfo {author} {\bibfnamefont {G.}~\bibnamefont {Gori}},
  \bibinfo {author} {\bibfnamefont {R.}~\bibnamefont {Vaia}}, \ and\ \bibinfo
  {author} {\bibfnamefont {P.}~\bibnamefont {Verrucchi}},\ }\href {\doibase
  10.1063/1.2172209} {\bibfield  {journal} {\bibinfo  {journal} {Journal of
  Applied Physics}\ }\textbf {\bibinfo {volume} {99}},\ \bibinfo {pages}
  {08H503} (\bibinfo {year} {2006})}\BibitemShut {NoStop}%
\bibitem [{\citenamefont {Ju}\ \emph {et~al.}(2012)\citenamefont {Ju},
  \citenamefont {Kallin}, \citenamefont {Fendley}, \citenamefont {Hastings},\
  and\ \citenamefont {Melko}}]{Ju2012}%
  \BibitemOpen
  \bibfield  {author} {\bibinfo {author} {\bibfnamefont {H.}~\bibnamefont
  {Ju}}, \bibinfo {author} {\bibfnamefont {A.~B.}\ \bibnamefont {Kallin}},
  \bibinfo {author} {\bibfnamefont {P.}~\bibnamefont {Fendley}}, \bibinfo
  {author} {\bibfnamefont {M.~B.}\ \bibnamefont {Hastings}}, \ and\ \bibinfo
  {author} {\bibfnamefont {R.~G.}\ \bibnamefont {Melko}},\ }\href {\doibase
  10.1103/PhysRevB.85.165121} {\bibfield  {journal} {\bibinfo  {journal} {Phys.
  Rev. B}\ }\textbf {\bibinfo {volume} {85}},\ \bibinfo {pages} {165121}
  (\bibinfo {year} {2012})}\BibitemShut {NoStop}%
\bibitem [{\citenamefont {Verresen}\ \emph {et~al.}(2018)\citenamefont
  {Verresen}, \citenamefont {Pollmann},\ and\ \citenamefont
  {Moessner}}]{Verresen2018}%
  \BibitemOpen
  \bibfield  {author} {\bibinfo {author} {\bibfnamefont {R.}~\bibnamefont
  {Verresen}}, \bibinfo {author} {\bibfnamefont {F.}~\bibnamefont {Pollmann}},
  \ and\ \bibinfo {author} {\bibfnamefont {R.}~\bibnamefont {Moessner}},\
  }\href {\doibase 10.1103/PhysRevB.98.155102} {\bibfield  {journal} {\bibinfo
  {journal} {Phys. Rev. B}\ }\textbf {\bibinfo {volume} {98}},\ \bibinfo
  {pages} {155102} (\bibinfo {year} {2018})}\BibitemShut {NoStop}%
\bibitem [{\citenamefont {Sar\'{i}yer}(2019)}]{Sariyer2019}%
  \BibitemOpen
  \bibfield  {author} {\bibinfo {author} {\bibfnamefont {O.~S.}\ \bibnamefont
  {Sar\'{i}yer}},\ }\href {\doibase 10.1080/14786435.2019.1605212} {\bibfield
  {journal} {\bibinfo  {journal} {Philosophical Magazine}\ }\textbf {\bibinfo
  {volume} {99}},\ \bibinfo {pages} {1787} (\bibinfo {year}
  {2019})}\BibitemShut {NoStop}%
\bibitem [{\citenamefont {Fisher}(1964)}]{fisher1964}%
  \BibitemOpen
  \bibfield  {author} {\bibinfo {author} {\bibfnamefont {M.~E.}\ \bibnamefont
  {Fisher}},\ }\href {\doibase 10.1119/1.1970340} {\bibfield  {journal}
  {\bibinfo  {journal} {American Journal of Physics}\ }\textbf {\bibinfo
  {volume} {32}},\ \bibinfo {pages} {343} (\bibinfo {year} {1964})}\BibitemShut
  {NoStop}%
\bibitem [{\citenamefont {Giamarchi}(2004)}]{giamarchi2004}%
  \BibitemOpen
  \bibfield  {author} {\bibinfo {author} {\bibfnamefont {T.}~\bibnamefont
  {Giamarchi}},\ }\href {\doibase 10.1093/acprof:oso/9780198525004.001.0001}
  {\emph {\bibinfo {title} {{Quantum physics in one dimension}}}},\
  International series of monographs on physics\ (\bibinfo  {publisher}
  {Clarendon Press},\ \bibinfo {address} {Oxford},\ \bibinfo {year}
  {2004})\BibitemShut {NoStop}%
\bibitem [{\citenamefont {Mila}(2000)}]{Mila_2000}%
  \BibitemOpen
  \bibfield  {author} {\bibinfo {author} {\bibfnamefont {F.}~\bibnamefont
  {Mila}},\ }\href {\doibase 10.1088/0143-0807/21/6/302} {\bibfield  {journal}
  {\bibinfo  {journal} {European Journal of Physics}\ }\textbf {\bibinfo
  {volume} {21}},\ \bibinfo {pages} {499} (\bibinfo {year} {2000})}\BibitemShut
  {NoStop}%
\bibitem [{\citenamefont {Franchini}(2017)}]{Franchini2017}%
  \BibitemOpen
  \bibfield  {author} {\bibinfo {author} {\bibfnamefont {F.}~\bibnamefont
  {Franchini}},\ }\href {\doibase 10.1007/978-3-319-48487-7} {\emph {\bibinfo
  {title} {{An Introduction to Integrable Techniques for One-Dimensional
  Quantum Systems}}}},\ Lecture Notes in Physics\ (\bibinfo  {publisher}
  {Springer Cham},\ \bibinfo {address} {Switzerland},\ \bibinfo {year}
  {2017})\BibitemShut {NoStop}%
\bibitem [{\citenamefont {Totsuka}(1998)}]{totsuka1998}%
  \BibitemOpen
  \bibfield  {author} {\bibinfo {author} {\bibfnamefont {K.}~\bibnamefont
  {Totsuka}},\ }\href {\doibase 10.1103/PhysRevB.57.3454} {\bibfield  {journal}
  {\bibinfo  {journal} {Phys. Rev. B}\ }\textbf {\bibinfo {volume} {57}},\
  \bibinfo {pages} {3454} (\bibinfo {year} {1998})}\BibitemShut {NoStop}%
\bibitem [{\citenamefont {Tonegawa}\ \emph {et~al.}(1998)\citenamefont
  {Tonegawa}, \citenamefont {Nishida},\ and\ \citenamefont
  {Kaburagi}}]{Tonegawa1998}%
  \BibitemOpen
  \bibfield  {author} {\bibinfo {author} {\bibfnamefont {T.}~\bibnamefont
  {Tonegawa}}, \bibinfo {author} {\bibfnamefont {T.}~\bibnamefont {Nishida}}, \
  and\ \bibinfo {author} {\bibfnamefont {M.}~\bibnamefont {Kaburagi}},\ }\href
  {\doibase https://doi.org/10.1016/S0921-4526(97)00937-X} {\bibfield
  {journal} {\bibinfo  {journal} {Physica B: Condensed Matter}\ }\textbf
  {\bibinfo {volume} {246-247}},\ \bibinfo {pages} {368} (\bibinfo {year}
  {1998})}\BibitemShut {NoStop}%
\bibitem [{\citenamefont {Mila}(1998)}]{Mila1998}%
  \BibitemOpen
  \bibfield  {author} {\bibinfo {author} {\bibfnamefont {F.}~\bibnamefont
  {Mila}},\ }\href {\doibase 10.1007/s100510050542} {\bibfield  {journal}
  {\bibinfo  {journal} {The European Physical Journal B - Condensed Matter and
  Complex Systems}\ }\textbf {\bibinfo {volume} {6}},\ \bibinfo {pages} {201}
  (\bibinfo {year} {1998})}\BibitemShut {NoStop}%
\bibitem [{\citenamefont {Chaboussant}\ \emph {et~al.}(1998)\citenamefont
  {Chaboussant}, \citenamefont {Julien}, \citenamefont {Fagot-Revurat},
  \citenamefont {Hanson}, \citenamefont {L{\'e}vy}, \citenamefont {Berthier},
  \citenamefont {Horvatic},\ and\ \citenamefont {Piovesana}}]{Chaboussant1998}%
  \BibitemOpen
  \bibfield  {author} {\bibinfo {author} {\bibfnamefont {G.}~\bibnamefont
  {Chaboussant}}, \bibinfo {author} {\bibfnamefont {M.~H.}\ \bibnamefont
  {Julien}}, \bibinfo {author} {\bibfnamefont {Y.}~\bibnamefont
  {Fagot-Revurat}}, \bibinfo {author} {\bibfnamefont {M.}~\bibnamefont
  {Hanson}}, \bibinfo {author} {\bibfnamefont {L.~P.}\ \bibnamefont
  {L{\'e}vy}}, \bibinfo {author} {\bibfnamefont {C.}~\bibnamefont {Berthier}},
  \bibinfo {author} {\bibfnamefont {M.}~\bibnamefont {Horvatic}}, \ and\
  \bibinfo {author} {\bibfnamefont {O.}~\bibnamefont {Piovesana}},\ }\href
  {\doibase 10.1007/s100510050539} {\bibfield  {journal} {\bibinfo  {journal}
  {The European Physical Journal B - Condensed Matter and Complex Systems}\
  }\textbf {\bibinfo {volume} {6}},\ \bibinfo {pages} {167} (\bibinfo {year}
  {1998})}\BibitemShut {NoStop}%
\bibitem [{\citenamefont {Tribedi}\ and\ \citenamefont
  {Bose}(2009)}]{Tribedi2009}%
  \BibitemOpen
  \bibfield  {author} {\bibinfo {author} {\bibfnamefont {A.}~\bibnamefont
  {Tribedi}}\ and\ \bibinfo {author} {\bibfnamefont {I.}~\bibnamefont {Bose}},\
  }\href {\doibase 10.1103/PhysRevA.79.012331} {\bibfield  {journal} {\bibinfo
  {journal} {Phys. Rev. A}\ }\textbf {\bibinfo {volume} {79}},\ \bibinfo
  {pages} {012331} (\bibinfo {year} {2009})}\BibitemShut {NoStop}%
\bibitem [{\citenamefont {Kawano}\ and\ \citenamefont
  {Takahashi}(1997)}]{kawano1997}%
  \BibitemOpen
  \bibfield  {author} {\bibinfo {author} {\bibfnamefont {K.}~\bibnamefont
  {Kawano}}\ and\ \bibinfo {author} {\bibfnamefont {M.}~\bibnamefont
  {Takahashi}},\ }\href {\doibase 10.1143/JPSJ.66.4001} {\bibfield  {journal}
  {\bibinfo  {journal} {Journal of the Physical Society of Japan}\ }\textbf
  {\bibinfo {volume} {66}},\ \bibinfo {pages} {4001} (\bibinfo {year}
  {1997})}\BibitemShut {NoStop}%
\bibitem [{\citenamefont {Tandon}\ \emph {et~al.}(1999)\citenamefont {Tandon},
  \citenamefont {Lal}, \citenamefont {Pati}, \citenamefont {Ramasesha},\ and\
  \citenamefont {Sen}}]{Tandon1999}%
  \BibitemOpen
  \bibfield  {author} {\bibinfo {author} {\bibfnamefont {K.}~\bibnamefont
  {Tandon}}, \bibinfo {author} {\bibfnamefont {S.}~\bibnamefont {Lal}},
  \bibinfo {author} {\bibfnamefont {S.~K.}\ \bibnamefont {Pati}}, \bibinfo
  {author} {\bibfnamefont {S.}~\bibnamefont {Ramasesha}}, \ and\ \bibinfo
  {author} {\bibfnamefont {D.}~\bibnamefont {Sen}},\ }\href {\doibase
  10.1103/PhysRevB.59.396} {\bibfield  {journal} {\bibinfo  {journal} {Phys.
  Rev. B}\ }\textbf {\bibinfo {volume} {59}},\ \bibinfo {pages} {396} (\bibinfo
  {year} {1999})}\BibitemShut {NoStop}%
\bibitem [{\citenamefont {Pushpan}\ \emph {et~al.}(2023)\citenamefont
  {Pushpan}, \citenamefont {KJ}, \citenamefont {Narayan},\ and\ \citenamefont
  {Pal}}]{Pushpan2023}%
  \BibitemOpen
  \bibfield  {author} {\bibinfo {author} {\bibfnamefont {C.~B.}\ \bibnamefont
  {Pushpan}}, \bibinfo {author} {\bibfnamefont {H.}~\bibnamefont {KJ}},
  \bibinfo {author} {\bibfnamefont {P.}~\bibnamefont {Narayan}}, \ and\
  \bibinfo {author} {\bibfnamefont {A.~K.}\ \bibnamefont {Pal}},\ }\href
  {\doibase 10.48550/arXiv.2301.04615} {\bibfield  {journal} {\bibinfo
  {journal} {arXiv:2301.04615}\ } (\bibinfo {year} {2023}),\
  10.48550/arXiv.2301.04615}\BibitemShut {NoStop}%
\bibitem [{\citenamefont {Sen(De)}\ and\ \citenamefont
  {Sen}(2010)}]{Sende2010}%
  \BibitemOpen
  \bibfield  {author} {\bibinfo {author} {\bibfnamefont {A.}~\bibnamefont
  {Sen(De)}}\ and\ \bibinfo {author} {\bibfnamefont {U.}~\bibnamefont {Sen}},\
  }\href {\doibase 10.1103/PhysRevA.81.012308} {\bibfield  {journal} {\bibinfo
  {journal} {Phys. Rev. A}\ }\textbf {\bibinfo {volume} {81}},\ \bibinfo
  {pages} {012308} (\bibinfo {year} {2010})}\BibitemShut {NoStop}%
\bibitem [{\citenamefont {Sadhukhan}\ \emph {et~al.}(2017)\citenamefont
  {Sadhukhan}, \citenamefont {Roy}, \citenamefont {Pal}, \citenamefont
  {Rakshit}, \citenamefont {Sen(De)},\ and\ \citenamefont
  {Sen}}]{Sadhukhan2017}%
  \BibitemOpen
  \bibfield  {author} {\bibinfo {author} {\bibfnamefont {D.}~\bibnamefont
  {Sadhukhan}}, \bibinfo {author} {\bibfnamefont {S.~S.}\ \bibnamefont {Roy}},
  \bibinfo {author} {\bibfnamefont {A.~K.}\ \bibnamefont {Pal}}, \bibinfo
  {author} {\bibfnamefont {D.}~\bibnamefont {Rakshit}}, \bibinfo {author}
  {\bibfnamefont {A.}~\bibnamefont {Sen(De)}}, \ and\ \bibinfo {author}
  {\bibfnamefont {U.}~\bibnamefont {Sen}},\ }\href {\doibase
  10.1103/PhysRevA.95.022301} {\bibfield  {journal} {\bibinfo  {journal} {Phys.
  Rev. A}\ }\textbf {\bibinfo {volume} {95}},\ \bibinfo {pages} {022301}
  (\bibinfo {year} {2017})}\BibitemShut {NoStop}%
\bibitem [{ggm()}]{ggm_definition}%
  \BibitemOpen
  \href@noop {} {}\bibinfo {note} {The generalized geometric measure for a
  multi-qubit pure state $\ket{\psi}$ is given by $G=1-\max\{\lambda^2\}$,
  where $\lambda$ is the maximum Schmidt coefficient of $\ket{\psi}$ for a
  specific bipartition, and the maximization in the expression for $G$ is
  performed over all arbitrary bipartitions of the multi-qubit
  system.}\BibitemShut {Stop}%
\bibitem [{\citenamefont {Mila}\ and\ \citenamefont
  {Schmidt}(2011)}]{Mila2011}%
  \BibitemOpen
  \bibfield  {author} {\bibinfo {author} {\bibfnamefont {F.}~\bibnamefont
  {Mila}}\ and\ \bibinfo {author} {\bibfnamefont {K.~P.}\ \bibnamefont
  {Schmidt}},\ }\enquote {\bibinfo {title} {Strong-coupling expansion and
  effective hamiltonians},}\ in\ \href {\doibase 10.1007/978-3-642-10589-0_20}
  {\emph {\bibinfo {booktitle} {Introduction to Frustrated Magnetism:
  Materials, Experiments, Theory}}},\ \bibinfo {editor} {edited by\ \bibinfo
  {editor} {\bibfnamefont {C.}~\bibnamefont {Lacroix}}, \bibinfo {editor}
  {\bibfnamefont {P.}~\bibnamefont {Mendels}}, \ and\ \bibinfo {editor}
  {\bibfnamefont {F.}~\bibnamefont {Mila}}}\ (\bibinfo  {publisher} {Springer
  Berlin Heidelberg},\ \bibinfo {address} {Berlin, Heidelberg},\ \bibinfo
  {year} {2011})\ pp.\ \bibinfo {pages} {537--559}\BibitemShut {NoStop}%
\end{thebibliography}%
 
\end{document}